\documentclass[12pt, oneside]{article} 

\usepackage{amsmath}  				
\usepackage{geometry}
\usepackage{amssymb}
\usepackage{mathtools}
\usepackage{slashed}
\usepackage{braket}
\usepackage[parfill]{parskip}    		
\usepackage{graphicx}
\usepackage{mathtools}

\allowdisplaybreaks
\usepackage[justification=centering]{caption}
\usepackage{mathtools}
\numberwithin{equation}{section}

\usepackage{xfrac}
\usepackage{amssymb}
\usepackage[table]{xcolor}
\usepackage{booktabs}
\usepackage{mdwtab}
\usepackage{breqn}
\usepackage{amsthm}
\theoremstyle{definition}
\usepackage{lmodern}
\def\a{\alpha}

\def\d{\delta}
\def\e{\epsilon}

\def\l{\lambda}
\def\m{\mu}
\def\n{\nu}
\def\o{\omega}

\def\r{\rho}
\def\s{\sigma}

\def\x{\xi}

\def\G{\Gamma}

\def\L{\Lambda}

\usepackage{float}
\usepackage{xcolor}
\usepackage{lscape}
\usepackage{subfig}
\usepackage{amsthm}
\usepackage{feynmp-auto}
\usepackage{graphicx}
\usepackage{caption}
\usepackage{tikz}
\usetikzlibrary{arrows.meta, decorations.markings, decorations.pathmorphing}

\def \nn {\nonumber}
\def \be  {\begin{equation}}
\def \ee  {\end{equation}}
\def \bea  {\begin{eqnarray}}
\def \eea  {\end{eqnarray}}

\newcommand{\ImCorr}[1]{%
  \mathcal{I}_{#1}}

\begin{document}
\begin{flushright}
		BONN-TH-2026-07
	\end{flushright}
	\begin{center}
		\Large \bf Spacelike and timelike structure functions: a dispersive crossing relation
	\end{center}
	
	\bigskip
	
	\centerline{ Aniruddha Venkata$^a$}
	\begin{center}
		${}^a$Bethe Center for Theoretical Physics, Universit$\ddot{\text{a}}$t Bonn, D-53115, Germany
	\end{center}
	\date{}

\abstract{Crossing symmetry suggests that deep inelastic scattering and semi inclusive electron–positron annihilation are governed by analytic continuations of a single forward amplitude. Drell, Levy, and Yan proposed that the hadronic tensor admits analytic continuation and demonstrated, in reasonable models, that connected contributions to the cross-section continue. They also identified obstructions to continuation of the current correlator. In this work we supplement their observation with a new dispersive proposal for analytic continuation of the correlator and, assuming polynomial boundedness, derive subtracted dispersion relations relating spacelike and timelike cross sections. We introduce a new factorized function that quantifies the obstruction to crossing and compute its hard kernel at lowest order. The resulting identity connects distribution functions in deep inelastic scattering to fragmentation functions in annihilation. 
}
\section{Introduction}
Crossing symmetry stands as one of the fundamental principles of quantum field theory, relating seemingly distinct scattering processes through analytic continuation of Lorentz invariants \cite{Eden:1966dnq}. First proposed by Gell-Mann, Goldberger, and Thirring \cite{Gell-Mann:1954ttj,Gell-Mann:1954wra}, crossing symmetry asserts that processes related by interchange of particles between initial and final states are described by a single analytic function evaluated in different kinematic regions. This principle is connected to causality and Lorentz invariance.

A rigorous foundation for crossing symmetry in local quantum field theory was established by Bros, Epstein, and Glaser \cite{Bros:1964iho,Bros:1965kbd}. Starting from the analyticity properties of Wightmann functions, they demonstrated that S-matrix elements are boundary values of analytic functions in complexified momentum space, with singularities determined by physical thresholds and crossed-channel cuts. Crossing symmetry thus emerges as a structural consequence of locality, unitarity, and Lorentz invariance.

In quantum electrodynamics, crossing symmetry is manifest order by order in perturbation theory. Classic examples relate Compton scattering to pair production and annihilation through analytic continuation of Mandelstam invariants \cite{Martin:1969ina}. The extraordinary agreement between QED predictions and precision measurements --- for example in determinations of the muon anomalous magnetic moment \cite{Aoyama:2020ynm} --- provides indirect confirmation of this analytic structure.

In quantum chromodynamics, the situation is more subtle. While the QCD Lagrangian respects the same fundamental principles of locality and Lorentz invariance, observable hadronic quantities involve long-distance matrix elements and infrared finite, perturbatively calculable hard functions. As a result, the analytic structure implied by crossing symmetry is not manifest at the level of individual contributions to measurable cross sections. Lattice calculations, typically formulated in Euclidean signature, do not directly impose the full Lorentzian analyticity properties underlying crossing symmetry. The implications of crossing symmetry for inclusive hadronic observables therefore require additional analysis.

Deep inelastic scattering (DIS) experiments determine spacelike structure functions with high precision, while electron--positron colliders measure hadron production through semi-inclusive annihilation (SIA) in the timelike region. These processes are related by crossing symmetry. Drell, Levy, and Yan (DLY) \cite{Drell:1969jm},\cite{Drell:1969wd},\cite{Drell:1969wb} observed that connected contributions to inclusive DIS and semi-inclusive annihilation structure functions arise as analytic continuations of each other. However, it was noted in \cite{Drell:1969wb} that a generic current commutator admits partially disconnected cuts which do not contribute to the SIA cross section but are non-vanishing in the SIA kinematic region. While the connected graphs admit analytic continuation, these disconnected contributions generate additional singularities that obstruct a naive identification of spacelike and timelike observables.

The extraneous contributions in the SIA region were later characterized in terms of a possible double discontinuity in \cite{Gatto:1972gpd},\cite{Gatto:1972ip},\cite{Dahmen:1973mn}, although it remained unclear whether such a double discontinuity is generically non-vanishing. The most general singularity structure of the current commutator was further analyzed using perturbative methods in \cite{Landshoff:1970ff},\cite{Suri:1971dc}. More recently, perturbative QCD calculations have shown that scheme-independent combinations of hard and collinear functions respect the DLY relation \cite{Blumlein:2000wh},\cite{Blumlein:2006rr}.

In this work we take the DLY observation as a starting point and, assuming polynomial boundedness and cut-plane analyticity, derive an exact dispersive crossing identity relating inclusive DIS structure functions to inclusive annihilation structure functions entirely in terms of physical cross sections. The identity contains a residual term $\Gamma_a$ arising from the partially disconnected cuts identified by DLY. A central result of this paper is that $\Gamma_a$ factorizes at leading power with the same parton distribution functions that enter DIS — introducing no new non-perturbative input — with a new, perturbatively computable hard coefficient. This factorization, established in Sec.~\ref{sec:disconnected}, is what renders the crossing identity fully predictive.

A natural motivation for such a relation arises from QCD factorization \cite{Collins:1989gx}. At leading power in the hard scale, DIS structure functions factorize into perturbative coefficient functions convoluted with parton distribution functions (PDFs), while inclusive hadron production in annihilation factorizes into analogous coefficient functions convoluted with fragmentation functions (FFs). Since the short-distance coefficients are related by crossing, the dispersive relation derived here implies, under leading-power collinear factorization, a corresponding connection between PDFs and FFs up to power-suppressed corrections.

This perspective clarifies the relation to the work of Gribov and Lipatov \cite{Gribov:1971zn},\cite{Gribov:1972rt}. Their analysis focused on perturbative evolution equations and identified a reciprocity between spacelike and timelike splitting functions at leading order. The Gribov-Lipatov relations were upgraded to non-perturbative ansatz for the crossing relationships \cite{Eylon:1974kw} and tested phenomenologically
\cite{Konoplyanikov:1993nf},\cite{Petrov:1999vx}. However, the Gribov-Lipatov relations were found to undershoot the data. More recent refinements of the Gribov-Lipatov relations (upgraded to non-perturbative relationships) have appeared in the literature \cite{Barone:2000tx} and have been phenomenologically tested \cite{Ma:2002ur} with success. In the framework developed here, Gribov–Lipatov like reciprocity relations arise as the leading-power, factorized consequence of Drell–Levy–Yan analyticity implemented through dispersion relations.

The plan of this paper is as follows. In Sec.~\ref{sec:results} we set up the kinematics and derive the dispersive crossing identity — the first central result of this paper — relating spacelike and timelike structure functions exactly, up to the residual term $\Gamma_a$. In Sec.~\ref{sec:field-theory} we state the analyticity and boundedness assumptions explicitly and derive, at leading power in the parton model, the Mellin-moment identity connecting parton distribution and fragmentation function moments — the second central result. In Sec.~\ref{sec:disconnected} we establish that $\Gamma_a$ factorizes with no new non-perturbative input, which is the ingredient that makes both results fully predictive. We conclude in Sec.~\ref{sec:conclusions}. The proof that the analyticity assumptions are satisfied to all orders in perturbation theory is given in Appendix~\ref{app:PT-analyticity}.

\section{Kinematics and Dispersion}
\label{sec:results}
Let us review the relevant kinematics in the context of DIS and $e^+ e^-$ experiments. Crossing symmetry relating the two experiments was first observed in a seminal paper by Drell, Levy and Yan  (henceforth DLY) \cite{Drell:1969jm}. Crossing symmetry is the statement that a single analytic function, continued to different kinematic regimes, describes disparate processes. In the particular situation of interest, we have
\bea
e^{-}(k)+e^{+}(k_1)\rightarrow \bar{p}(\bar{P})+X,\nn \\
e^{-}(k)+ p(P) \rightarrow e^{-}(k_2) + X.
\eea
In $e^{+}e^{-}$ it is conventional to define the kinematic variable $q=k+k_1$, and observe that $q^2 >0$ since onshell timelike hadrons are produced through pair annihilation. Furthermore $\bar{P}$ is the momentum of a timelike hadron and we have $q\cdot \bar{P} \geq 0$. 
We symbolize the amplitude for the process by $\mathcal{M}_{ep}^{s_e,s_p,s_{\bar{P}}}(q^2,q\cdot \bar{P},\{P_X\})$. Here, the superscripts are spin indices and the subscript indicates that we are describing electron positron annihilation. In what follows, we assume that the beams are unpolarized and the spin of the antiproton is unobserved. We suppress spin indices and sum over all spins in the final state and average over the spins in the initial state when constructing cross sections.

In DIS, the kinematic variables are $Q^2=(k-k_2)^2, Q\cdot P$. Say the amplitude  for this process is $\mathcal{M}_{\text{DIS}}(Q^2,Q\cdot {P},\{P_X\})$

In this case, because $Q$ is the momentum transferred between two identical, on-shell particles, it is  a space-like vector.
The second independent Mandelstam invariant, $Q\cdot P$ is positive. We may infer this by requiring that the produced particle have timelike momentum, that is $P_X$ is time-like.
\bea
P_X^2=(P+Q)^2=m_P^2+2P\cdot Q (1-x_B),
\eea
where we have introduced the Bjorken scaling variable $x_B=\frac{-Q^2}{2P\cdot Q}$ and the proton mass $m_P$. The Bjorken scaling variable, $x_B$ lies between $(0,1)$ in the $m_P \rightarrow 0$ limit.   So, in DIS, the region of interest is $Q^2<0,-P\cdot Q < 0$. 
Crossing is the statement that these two processes, related by particles moving from in state to the out state (or vice-versa), are described by the same analytic function, albeit in disjoint regions of parameter space.  
\bea
\mathcal{M}_{\text{DIS}}(Q^2,Q\cdot {P},\{P_X\})=-\mathcal{M}_{ep}(Q^2,-Q\cdot {P},\{P_X\}).\label{eq:crossing-ME}
\eea
Although they are the same analytic function, they evaluate to different values because they are evaluated in disjoint regions. Therefore, a direct comparison of data is not possible.
 In their seminal paper, DLY conclude that the inclusive cross-section itself is an analytic function of the Mandelstam invariants. In general, cross sections are not analytic because the modulus squared of an analytic function is manifestly non holomorphic. However, for inclusive cross-sections, DLY argued that the cross-section level observables are themselves analytic functions since they can be written as the imaginary part of a matrix element which is analytic.  

We now describe how to directly analytically continue the functions from the electron positron region to the DIS region, and compare cross sections.
The DIS cross section is understood to be described by hadronic tensors $W_{1,2}(Q^2,x_B)$. Let us define the hadronic tensors by relating it to the cross section
\bea
\frac{|
\vec{k}_2| d\sigma}{d^3k_2}= L_{\mu \nu}(k,k_2,s,-Q^2) W^{\m\n}(P,Q),
\eea
where the leptonic tensor contains leptonic factors, the photon propagator terms and other constants of proportionality between the cross section and the hadronic tensor. Explicitly, it is given by 
\begin{equation}
    L^{\mu\nu}(k, k_2,s,-Q^2) =\frac{e^2}{2s\left(4\pi^2 (-Q^2)^2\right)}
   \frac{1}{2} \,\mathrm{Tr}\!\left[\slashed{k}\,\Gamma^\mu\,\slashed{k}_2\,\Gamma^\nu\right],
    \label{eq:Lmunu}
\end{equation}

The hadronic tensor $W^{\m\nu}(P,Q)$, is described entirely by two functions $W_{1,2}(Q^2,2p\cdot Q)$, and the tensor structure is completely fixed by symmetry
\bea
W^{\mu\nu}&=&\left(\frac{Q^{\m}Q^{\n}}{Q^2}-\eta^{\m\n}\right)W_{1}(Q^2,2P\cdot Q)\nn \\ &&+\frac{1}{4x_B^2}\left(2x_B P^\m+Q^{\m}\right)\left(2x_B P^\n+Q^{\n}\right)\frac{W_2 (Q^2,2P\cdot Q)}{P\cdot Q}.\label{eq:hadronic_tensor}
\eea
Furthermore, $W_{1,2}$ are the imaginary parts of an  analytic function $T_{1,2}$. 
\bea
W_a(Q^2,2P\cdot Q)=2\text{Im}[T_a(Q^2,2P\cdot Q)]. \label{eq:Disc-DIS}
\eea
We remind the reader that the DIS region is $Q^2<0, 2P.Q>0$.  
It is known \cite{Sterman:1993hfp} that at fixed $Q^2<0$, the functions $T_a$ only have singularities for real values of $x_B$, in the interval $(-1,1)$ and satisfies the relationship
\bea
T_a(Q^2,x)=T_a(Q^2,-x), \hspace{2cm} Q^2<0.
\eea
We will interchangeably think of $T_a(Q^2,2P\cdot q)$, and $T_a(Q^2,x=\frac{-Q^2}{2P\cdot Q})$ as being the same analytic function, although they are formally distinct objects. 

\begin{figure}[h]
  \centering
  \includegraphics[width=0.8\textwidth]{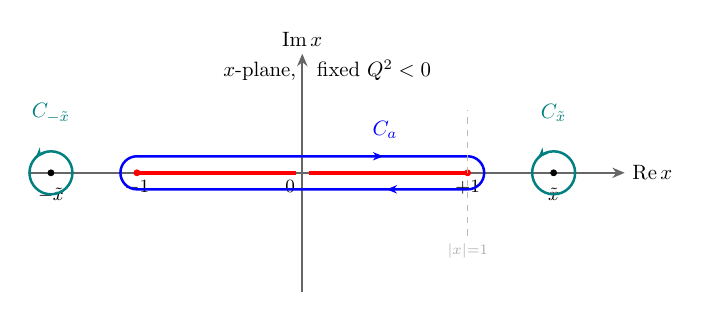}
  \caption{Analytic structure of $T_a(Q^2,x)$ in the complex $x$-plane
    at fixed $Q^2<0$. Branch cuts (red) lie on $x\in(-1,0)\cup(0,1)$.
    The contour $C_a$ (blue) encircles the cut, picking up
    $\mathrm{Disc}\,T_a = 2i\,W_a$. After deformation, $C_a$ collapses
    onto the residue contour $C_{\pm\tilde{x}}$ (teal) encircling the
    evaluation points $\tilde{x} = 1/x_F > 1$. }
  \label{fig:x-plane-contour}
\end{figure}

Therefore, in terms of the hadronic tensors $W_{1,2}(Q^2,x_B)$, measured in DIS experiments, we have a  dispersion relation
\bea
{T_{a}(Q^2, \tilde{x})}&=& \int_{C_a} \frac{dx}{2\pi i} \frac{xT_a(Q^2,x)}{x^2-\tilde{x}^2}, \nn \\ &=&\int_{0}^{1} \frac{xW_{a}(Q^2,x)}{x^2-\tilde{x}^2} \frac{dx}{\pi},
\eea
where we have defined the contour $C_a=(-1-i\e , 1+i\e )\cup (1-i\e ,-1-i\e) $ consisting of two line segments running in opposite directions and left hand side follows from deforming this contour to wrap around the positive real axis which has a pole contribution from $\pm \tilde{x}$. The imaginary part is an odd function which requires that we multiply the numerator by $x$ when carrying out the dispersion integral. The dispersion relation is depicted in Fig.\ \ref{fig:x-plane-contour}, displaying the contours explicitly.  Here, $\tilde{x}$ is subtraction point satisfying $|\tilde{x}|>1$. 

The observation we make here is that the left hand side of this equation can also be written in terms of electron positron annihilation  data, assuming crossing symmetry, up to a double discontinuity function which we will describe in Sec.\ref{sec:disconnected}. This gives us a way to compare the two experimentally accessible regions.  

Let us now turn to electron positron annihilation. First, we observe that the total $e^{+}e^{-}$ cross-section is the imaginary part of a analytic function
\bea
\sigma_{\text{tot}}=\bar{L}_{\m \n} (k,k_1,s,q^2) \bar{H}^{\m \nu} (q^2),
\eea
where the leptonic tensor contains lepton spinors, the overall factors of $\frac{1}{s}$, spin average factors, and other constants.
\bea
    \bar{L}^{\mu\nu}(k, k_1,s,q^2) &=& \frac{-e^2}{16\pi^2\,s (q^2)^2}
    \,\mathrm{Tr}\!\left[\slashed{k}\,\Gamma^\mu\,\slashed{k}_1\,\Gamma^\nu\right],\nn \\ 
    &=& {L}^{\mu\nu}(k, -k_1,s,q^2).
    \label{eq:Lbarmunu}
\eea
The hadronic tensor $\bar{H}^{\m \n}$ is the imaginary part of the photon self-energy
\bea
\bar{H}^{\m\n}(q^2)&=& 2\text{Im}\left[\pi^{\mu\nu}(q^2)\right], \nn \\ 
\pi^{\mu\nu}(q^2)&=& \left(\frac{q^{\m}q^{\n}}{q^2}-\eta^{\m\n}\right) \pi (q^2),
\eea
and $\pi(q^2)$ admits a spectral representation
\bea
\pi(q^2)= \int d\mu^2  \frac{\rho (\mu^2)}{q^2-\mu^2+i\e}.
\eea
We observe that the function $\pi(q^2)$ is only singular on the positive real axis, analytic continuation to negative $q^2$ can be done dispersively.  

When we observe a final state proton in this process, the leptonic part of the cross section is unchanged. However, we may define a hadronic tensor that now depends on both $\bar{P},q$. The most general tensor structure can be inferred by demanding transversality of the photon and invariance under parity as in DIS.
\bea
\frac{|\vec{\bar{P}}|d\s}{d^3{\bar{P}}}&=&\bar{L}_{\mu\nu}(q,k,k_1)\bar{W}^{\m\n}(\bar{P},q), \\
\bar{W}^{\m\n}(q,\bar{P})&=& \left(\frac{q^{\m}q^{\n}}{q^2}-\eta^{\m\n}\right)\bar{W}_{1}(q^2,2\bar{P}\cdot q)\nn \\ &&-\frac{x_F^2}{4}\left(\frac{2}{x_F}\bar{P}^\m-q^{\m}\right)\left(\frac{2}{x_F}\bar{P}^\n-q^{\n}\right)\frac{\bar{W}_2 (q^2,2 \bar{P}\cdot q)}{\bar{P}\cdot q},\nn
\eea
where we defined $x_F= \frac{2\bar{P}.q}{q^2}$ which lies between $(0,1)$ in the physical region.
We claim that crossing is equivalent to the assertion
\bea
\bar{W}_{a}(q^2, 2\bar{P}\cdot q)+\Gamma_a(q^2,2\bar{P}\cdot q)=2\text{Im}[T_a(q^2,-2\bar{P}\cdot q)].\label{eq:crossing213}
\eea
The ``residual'' function $\Gamma_a$ arises from partially disconnected cuts of the time-ordered product that are kinematically accessible in the SIA region but do not contribute to the SIA cross section. As established in Sec.~\ref{sec:disconnected}, $\Gamma_a$ factorizes at leading power with the same parton distribution functions as DIS, with a new perturbatively computable hard coefficient and no new non-perturbative input. This factorization is what makes the crossing identity fully predictive.
To  analytically continue $T_a$ to a region where $q^2$ is negative, the region where DIS  is naturally defined, we proceed dispersively as before. Consider the function $T_a(q^2,x=\frac{-q^2}{(-2\bar{P}\cdot q)})$. We fix $x=\tilde{x}=\frac{1}{x_F}>1$, and write a dispersion relation for the functions $T_{a}(q^2, x)$ in $q^2$. As proved in Appendix~\ref{app:PT-analyticity}, the function $T_a(q^2,\tilde{x})$ for fixed $\tilde{x}>1$ only has singularities on the real $q^2$ axis for positive values of $q^2$. As a result, we can dispersively 
continue the function to negative values of $q^2$.
\begin{figure}[ht]
  \centering
  \includegraphics[width=0.8\textwidth]{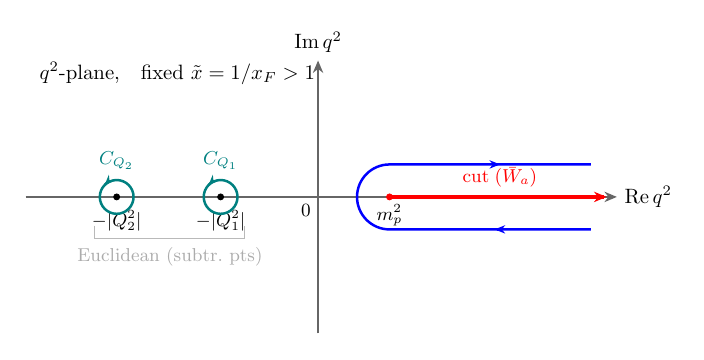}
  \caption{Analytic structure of $T_a(q^2,\tilde{x})$ in the complex
    $q^2$-plane at fixed $\tilde{x} = 1/x_F > 1$. The branch cut (red)
    lies on the positive real axis for $q^2 \ge m_p^2$, where
    $\mathrm{Disc}\,T_a = 2i\,\bar{W}_a+2i\Gamma_a$. The Hankel contour
    $\mathcal{C}$ (blue) wraps the cut; the arc at infinity 
    vanishes by polynomial boundedness, yielding
    Eq.~\eqref{eq:dispersion-SIA}. After deformation, $\mathcal{C}$
    collapses onto the residue contours $C_{Q_1}$ and $C_{Q_2}$ (teal)
    encircling the Euclidean subtraction points $-|Q_1^2|$ and
    $-|Q_2^2|$.}
  \label{fig:q2-plane-contour}
\end{figure}
The dispersion relation reads:
\bea
\frac{T_a\left(-|Q^2_1|, \tilde{x}\right)}{|Q_2^2|-|Q_1^2|}+\frac{T_a\left(-|Q^2_2|, \tilde{x}\right)}{|Q_1^2|-|Q_2^2|}&=& \int_{m_{p}^2} ^{\infty} \frac{\bar{W}_{a}(q^2,x_F=\frac{1}{\tilde{x}})}{(q^2+|Q_1^2|)(q^2+|Q_2^2|)}\frac{dq^2}{2\pi} \label{eq:dispersion-SIA} \\ &&+ \int_{0} ^{\infty}\frac{\G_{a}(q^2,x_F=\frac{1}{\tilde{x}})}{(q^2+|Q_1^2|)(q^2+|Q_2^2|)} \nn 
\eea
The contours appearing in this relation are shown in Fig.\ \ref{fig:q2-plane-contour}. 
The  right hand side of this dispersive integral is the experimentally accessible region of $0<x_F<1,q^2>0$. The left hand side is the function evaluated in a Euclidean region, an entirely unphysical region where no scattering occurs. Throughout this, we assume that two subtractions are sufficient to ensure that arcs at infinity don't contribute. The branch cut for the timelike DIS cross-section, which is equal to the residual function in the massless quark limit, starts at $0$ which explains why the dispersive integral for this function has a branch different point. 

Combining the DIS and SIA dispersion relations, we arrive at the first central result of this paper: an exact identity relating inclusive spacelike and timelike structure functions,
\bea
\frac{2}{(|Q_2^2|-|Q_1|^2)}\int_{0}^{1} \frac{xW_{1,2}(Q^2_1,x)}{x^2-\tilde{x}^2} \frac{dx}{\pi}+\frac{2}{(|Q_1^2|-|Q_2|^2)}\int_{0}^{1} \frac{xW_{1,2}(Q^2_2,x)}{x^2-\tilde{x}^2} \frac{dx}{\pi}\nn \\ =
\int_{0} ^{\infty} \frac{\theta(q^2-m_p^2)\bar{W}_{1,2}(q^2,x_F=\frac{1}{\tilde{x}})+\Gamma_{1,2}(q^2,x_F)}{(q^2+|Q_1^2|)(q^2+|Q_2^2|)}\frac{dq^2}{\pi}.\label{eq:result}
\eea

The left-hand side is built entirely from DIS structure functions, evaluated dispersively at two Euclidean subtraction points. The right-hand side is the SIA cross section integrated against the same dispersion kernel, plus the residual term $\Gamma_a$. As established in Sec.~\ref{sec:disconnected}, $\Gamma_a$ factorizes with no new non-perturbative input, so the identity is fully predictive up to power corrections. The physical content of this identity — and in particular the connection to parton distributions and fragmentation functions — is made explicit in Sec.~\ref{sec:field-theory}, where imposing leading-power factorization on both sides yields the second central result of this paper: a concrete Mellin-space identity relating PDF and fragmentation function moments.

Before we conclude this section, let us describe the notation used throughout this paper. We will reserve the symbol $x$ for the Bjorken scaling variable, which by definition is $x=-\frac{Q^2}{2P\cdot Q}$. We reserve the symbol $\o$ for the inverse momentum fraction variable $\o=\frac{1}{x}$. In the SIA region we will consistently use the symbol $x_F$ for the variable Feynman momentum fraction $\frac{2\bar{P}\cdot q}{q^2}$ and $\omega_F=\frac{1}{x_F}$.  

In the next section, we state the assumptions underlying these dispersion relations and derive the leading-power factorized form of the identity using the parton model.
\section{The Structure Functions in Field Theory}
\label{sec:field-theory}

In this section, we provide a field-theoretic interpretation for the invariant amplitudes $T_a$ and justify the dispersive framework established in Sec.~\ref{sec:results}. We begin by stating the assumptions regarding the analytic structure and asymptotic behavior of the hadronic tensor; these assumptions are verified to all orders in perturbation theory in Appendix~\ref{app:PT-analyticity}. We then impose leading-power collinear factorization on both sides of the dispersion relation and derive, within the parton model, an explicit identity relating the Mellin moments of parton distribution functions to those of fragmentation functions.

Within a field theory setup, the invariant amplitudes $T_a$ are obtained by projecting the forward virtual Compton tensor,
\begin{equation}
    T^{\mu\nu}(P, q) = \frac{i}{8\pi} \int d^4 x\, e^{iq \cdot x} 
    \bra{P} \mathcal{T}\left\{ J^\mu(x) J^\nu(0) \right\} \ket{P},\label{eq:correlator}
\end{equation}
onto the scalar structures defined in Eq.~\eqref{eq:hadronic_tensor}. 
Explicitly,
\bea
T^{\mu\nu}(P, Q)&=&\left(\frac{Q^{\m}Q^{\n}}{Q^2}-\eta^{\m\n}\right)T_{1}(Q^2,2P\cdot Q)\nn \\ &&+\frac{1}{4x_B^2}\left(2x_B P^\m+Q^{\m}\right)\left(2x_B P^\n+Q^{\n}\right)\frac{T_2 (Q^2,2P\cdot Q)}{P\cdot Q}.
\eea
In a spectral representation $T_a$ can be expressed as a sum over intermediate states $X$,
\bea
T^{\mu\nu}(P, q) &=& \frac{i}{8\pi} \int d^4 x\, e^{iq \cdot x}  \sum_{X}\theta(x^0) \Bigg (\langle P| J^{\m}(x)|X\rangle \langle X|J^{\n}(0)|P\rangle\nn \\ &&+\theta(-x^0)\langle P| J^{\m}(0)|X\rangle \langle X|J^{\n}(X)|P\rangle\Bigg).
\eea
We may utilize the integral representation of the $\theta$ function,
\bea
\theta(\pm x)= \mp \int\frac{d\o}{2\pi i} \frac{e^{-i\o x}}{\o\pm i\e},
\eea
and the translation operator to write
\bea
T^{\mu\nu}(P, q) &=& \frac{(2\pi)^3}{8\pi}  \sum_{X} \Bigg (\frac{\langle P| J^{\m}(0)|X\rangle \langle X|J^{\n}(0)|P\rangle}{q^0+P^0-P_X^0+i\e} \d^3\left(\vec{q}+\vec{P}-\vec{P}_X\right)\nn \\ &&+\frac{\langle P| J^{\m}(0)|X\rangle \langle X|J^{\n}(0)|P\rangle}{q^0-P^0+P_X^0-i\e}\Bigg).
\eea
In the DIS region, the imaginary part conserves all four components of momentum and satisfies, $(Q+P)^2=P_X^2=2P\cdot Q (1-x)$ which in turn implies that $0<x<1$. On the other hand the second term satisfies $(Q-P)^2=P_X^2=-2P\cdot Q+Q^2$ which can never be satisfies for timelike final states in the $Q^2<0,2P\cdot Q >0$ region. As a result only the first term contributes to the imaginary part. The imaginary part of this term yields the structure functions $W_a$ as in Eq.\ (\ref{eq:Disc-DIS}), while in the timelike region, they relate to the fragmentation functions $\bar{W}_a+\Gamma_a$ via the crossing relation in Eq.~\eqref{eq:crossing213} as we will discuss. 

\subsection{Assumptions for dispersion relations}
The derivation of the dispersive relation in Sec.~\ref{sec:results} rests upon three primary assumptions concerning the forward virtual Compton amplitude.

\begin{enumerate}
    \item \textbf{Cut Plane Analyticity:} We assume that for a fixed $Q^2$, the amplitudes $T_a(Q^2, x)$ are analytic functions in the complex $x$-plane, with singularities restricted to the real axis. Specifically, the amplitudes possess branch cuts on the real intervals $x \in (-1, 1)$ corresponding to the physical DIS and SIA regions. This cut plane analyticity is implicit in the construction of the dispersion relations in Sec.~\ref{sec:results} and, as proved in Appendix~\ref{app:PT-analyticity}, is satisfied to all loop orders in perturbation theory.
    
    \item \textbf{Crossing Symmetry of Matrix Elements:} We assume that the underlying matrix elements $\mathcal{M}$ satisfy the crossing symmetry relation established in Eq.~\eqref{eq:crossing-ME}:
    \begin{equation}
        \mathcal{M}_{\text{DIS}}(Q^2, Q\cdot P, \{P_X\}) = -\mathcal{M}_{ep}(Q^2, -Q\cdot P, \{P_X\}).
    \end{equation}
    This assumption identifies the DIS and electron-positron annihilation processes as different kinematic realizations of the same analytic function.  This enables us to identify cuts which correspond to the process $\gamma^{*}\rightarrow\bar{P}+X$ and appear in the analytic continuation of $T_a$ to the production of an anti-proton.

    \item \textbf{Polynomial Boundedness:} To ensure the convergence of the twice-subtracted dispersion relations, we assume that the amplitudes $T_a$ are polynomially bounded at large values of their arguments. Specifically, we require that the growth in both the scaling variable $x$ (at fixed $Q^2$) and the hard scale $Q^2$ (at fixed $x$) is strictly less than quadratic:
    \begin{align}
        \lim_{|x| \to \infty} \frac{T_a(Q^2, x)}{x} &= 0, \label{eq:x-growth} \\
        \lim_{|Q^2| \to \infty} \frac{T_a(Q^2, x)}{(Q^2)^2} &= 0.
    \end{align}
    These conditions justify discarding the contributions from the integration arcs at infinity, allowing the amplitudes to be reconstructed entirely from their discontinuities (the physical structure functions $W_a$ and $\bar{W}_a$) and the subtraction point data. While it is standard to assume that two subtractions are sufficient for enforcing vanishing at the large values of the invariant \cite{Eden:1966dnq}, the fact that $T_a$ is an even function of $x$, along with the assumption of analyticity in $x$, justifies Eq.\ (\ref{eq:x-growth}).
\end{enumerate}

The analyticity properties listed in Assumption 1 are verified to all orders in perturbation theory in Appendix~\ref{app:PT-analyticity}, via an explicit analysis of the Feynman-parametric representation using the Symanzik polynomial structure.

\subsection{Consequences of Leading-Power Approximation}

The dispersive crossing identity Eq.~\eqref{eq:result} takes its most concrete and predictive form when leading-power collinear factorization is imposed on both sides. The result is the second central result of this paper: a Mellin-space identity expressing the moments of fragmentation functions entirely in terms of PDF moments and perturbatively computable dispersed hard coefficients, with no new non-perturbative input. We now derive this identity, using the factorization of $\Gamma_a$ established in Sec.~\ref{sec:disconnected}. When studying factorized expressions, it is natural to dispersively continue Mellin moments of the structure function rather than the structure function itself. In general, a convolution reduces to a product in Mellin space due to the identities
\bea
f(x)&=&\int_x^1 \frac{d\xi}{\xi} \; H\left(\frac{x}{\xi}\right) \; \phi (\xi),\nn \\
f_N&=& \int_0^1 dx \; x^{N-1}f(x),\nn \\ 
&=& H_N \; \phi _N,
\eea
where we assume that $\xi$ is restricted to the interval $(0,1)$. 
In what follows, we would like to relate the Mellin moments of the distribution functions to data determined from fragmentation functions. 

At leading power in $Q^2$, the DIS structure functions factorize as
\begin{equation}
  W_a(Q^2, x)
  \;=\;
  \sum_i \int_x^1 \frac{d\xi}{\xi}\,
  C_{a,i}\!\left(\frac{x}{\xi}, \frac{Q^2}{\mu_D^2}\right)
  f_i(\xi, \mu_D^2)
  \;+\; \mathcal{O}\!\left(\frac{1}{Q^2}\right),
  \label{eq:DIS_factorization}
\end{equation}
where $C_{a,i}$ are the perturbative hard-scattering coefficients,
$f_i(\xi,\mu_D^2)$ are the parton distribution functions at
factorization scale $\mu_D$, and the sum runs over parton species $i$. By choosing the subtraction points, $Q_1^2,Q_2^2$, where we will evaluate $W_a(Q^2,x)$ to be large, this approximation may be applied safely throughout the bulk of the dispersive $x$ region.  In Eq.\ (\ref{eq:DIS_factorization}) the parton's momentum fraction $\xi$ cannot be arbitrarily small because of the spectral constraint $(\xi P+Q)^2\geq 0$. 

Similarly, the factorization theorem for semi-inclusive  annihilation reads
\begin{equation}
  \bar{W}_a(q^2, x_F)
  \;=\;
  \sum_i \int_{x_F}^1 \frac{d\xi}{\xi}\,
  H_{a,i}\!\left(\frac{x_F}{\xi}, \frac{q^2}{\mu_F^2}\right)
  D_i(\xi, \mu_F^2)
  \;+\; \mathcal{O}\!\left(\frac{1}{Q^2}\right),
  \label{eq:SIA_factorization}
\end{equation}
where $H_{a,i}$ are the perturbative hard-scattering coefficients,
$D_i(\xi,\mu_F^2)$ are the fragmentation functions at
factorization scale $\mu_F$, and the sum runs over parton species $i$. Since SIA annihilation produces at least one proton-anti proton pair (in addition to other particles in the final state), the threshold lies at $4m_p^2\gg \L_{QCD}$, we expect this leading power approximation to be accurate throughout the $q^2$ region.  In this case, the hadron carries a fraction $\xi$ of the parton momentum, but the energy fraction of the parton kinematically cannot exceed the total available energy, which sets a lower limit on the parton's momentum.    

Finally, we propose that the residual function $\Gamma_a$ also factorizes, with no new data at long distances
\begin{equation}
  \Gamma_a\left(q^2, x_F=\frac{1}{\o_F}\right)
  \;=\;
  \sum_i \int_{0}^1 \frac{d\xi}{\xi}\,
  K_{a,i}\!\left(\frac{\o_F}{\xi}, \frac{q^2}{\mu_r^2}\right)
  f_i(\xi, \mu_r^2)
  \;+\; \mathcal{O}\!\left(\frac{1}{Q^2}\right),
  \label{eq:disc_factorization}
\end{equation}
where the short distance coefficients are perturbatively calculable objects. In Sec.\ \ref{sec:disconnected} we supply an algorithm to compute this function in perturbation theory.  One does not expect that the factorization theorem holds down to arbitrarily small $q^2$. However, just as in Sec.\ \ref{sec:results}, the dispersive integrals over the spectrum will be algebraically equal to evaluating the class of graphs that contribute to $\Gamma_a$ at large negative $Q_i^2$, the chosen subtraction points through analytic continuation. The OPE at large negative $Q_i^2$ and as a result, $\Gamma_a$ respects leading power factorization \cite{Sterman:1993hfp}. Therefore, one expects that the scales $Q_i^2$ control the deviations from the leading power approximation.   Here,  the kinematic spectral constraint does not exist, and the residual timelike DIS cross-sections receives contributions from a partons of arbitrarily small momentum fraction $\xi$.

Our starting point is the Mellin transform in $x_F$ of the right hand side of Eq. (\ref{eq:result}). Interchanging the order of the $x_F$-integral and the
$q^2$-integral, yielding directly the $N$-th Mellin
moment of the SIA structure function integrated against the double
dispersion kernel:
\be
  \int_0^1 dx_F\;x_F^{N-1}
  \!\int_{m_p^2}^{\infty}\!
  \frac{\bar{W}_a(q^2,x_F)}{(q^2+|Q_1^2|)(q^2+|Q_2^2|)}
  \frac{dq^2}{2\pi}
  \;=\;
  \int_{m_p^2}^{\infty}
  \frac{\widetilde{\bar{W}}_a(q^2,N)}{(q^2+|Q_1^2|)(q^2+|Q_2^2|)}
  \frac{dq^2}{2\pi},
  \label{eq:RHS_Mellin}
\ee
where
$\widetilde{\bar{W}}_a(q^2,N)\equiv
\int_0^1 dx_F\, x_F^{N-1}\,\bar{W}_a(q^2,x_F)$
is the $N$-th Mellin moment of the SIA structure function at fixed
$q^2>0$.

The right-hand side of~\eqref{eq:RHS_Mellin} reads
\be
  \int_{m_p^2}^{\infty}
  \frac{\widetilde{\bar{W}}_a(q^2,N)}{(q^2+|Q_1^2|)(q^2+|Q_2^2|)}
  \frac{dq^2}{2\pi}
  \;=\;
  \sum_i
  \widehat{H}_{a,i}(N,Q_1^2,Q_2^2,\mu^2)\,
  \tilde{D}_i(N,\mu^2),
  \label{eq:RHS_factorized}
\ee
where the dispersively continued SIA hard coefficient
\bea \widehat{H}_{a,i}(N,Q_1^2,Q_2^2,\mu^2)
 \equiv \int_{m_p^2}^{\infty}\frac{dq^2}{2\pi}
 \frac{\widetilde{\bar{H}}_{a,i}(N,q^2/\mu^2)}
      {(q^2+|Q_1^2|)(q^2+|Q_2^2|)},\eea 
is a calculable perturbative quantity.
We may apply the transform to the residual function as well, but it does not decouple the distribution from the hard coefficient because the lower limit on the momentum fraction does not exist. 
\bea
\int_0^1 dx_F\;x_F^{N-1}
  \!\int_{0}^{\infty}\!
  \frac{\Gamma_a(q^2,x_F)}{(q^2+|Q_1^2|)(q^2+|Q_2^2|)}
  \frac{dq^2}{2\pi}
  \;=\;\sum_i \widehat{K}_{a,i}(N,Q_1^2,Q_2^2,\m_r^2) \label{eq:residual-function-factorized},
\eea
where the dispersively continued residual function has been written in Mellin space 
\bea \widehat{K}_{a,i}(N,Q_1^2,Q_2^2,\mu^2)
 \equiv \int_0^1 \frac{d\xi}{\xi} \; \int_{0}^{\infty}\frac{dq^2}{2\pi}\int_0^1 dx_F \; x_F^{N-1} \nn \\ 
 \frac{{{K}}_{a,i}(x_F \xi,\;q^2,/\mu^2)}
      {(q^2+|Q_1^2|)(q^2+|Q_2^2|)}
      \times f_i(\xi,\mu^2).\label{eq:residual-mellin}
\eea
By Eq.\ (\ref{eq:dispersion-SIA}), the dispersively continued objects in Eq.\ (\ref{eq:RHS_factorized}) and Eq.\ (\ref{eq:residual-function-factorized}) are related to correlation function by
\bea
\int_0^{1} d\tilde{x} \; \; \tilde{x}^{N-1}\left(\frac{T_a\left(-|Q^2_1|, \tilde{x}\right)}{|Q_2^2|-|Q_1^2|}+\frac{T_a\left(-|Q^2_2|, \tilde{x}\right)}{|Q_1^2|-|Q_2^2|}\right)&=&\nn \\ && \hspace{-8cm}\sum_i
  \widehat{K}_{a,i}(N,Q_1^2,Q_2^2,\mu^2)\,
  +\sum_i
  \widehat{H}_{a,i}(N,Q_1^2,Q_2^2,\mu^2)\,
  \tilde{D}_i(N,\mu^2). \label{eq:final-fact-SIA}
\eea

\begin{figure}[h]
  \centering
  \includegraphics[width=0.8\textwidth]{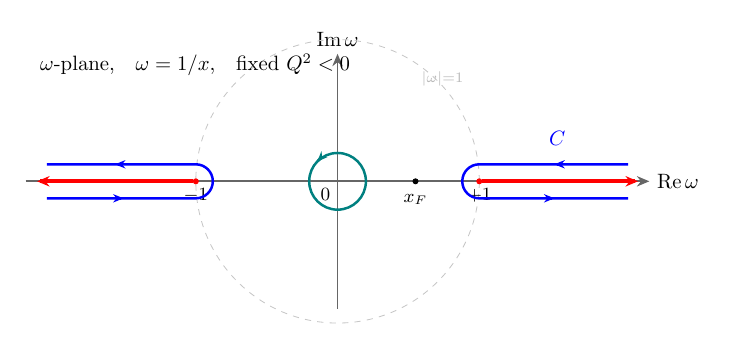}
  \caption{Analytic structure of $T_a(Q^2,\omega)$ in the complex
    $\omega = 1/x$ plane at fixed $Q^2<0$. Branch cuts (red) lie on
    the real axis for $|\omega|\ge 1$, i.e.\
    $\omega\in(-\infty,-1]\cup[1,\infty)$, with the unit circle
    $|\omega|=1$ shown as a dashed reference. The contour $C$ (blue)
    wraps each semi-infinite cut, picking up
    $\mathrm{Disc}\,T_a = 2i\,W_a$. Inside the unit disk $T_a$ is
    analytic and admits the Taylor expansion
    $T_a = \sum_N c_N\,\omega^N$. The teal contour encircles
    $\omega=0$ and extracts the Taylor coefficients via the Cauchy
    formula; the point $x_F\in(0,1)$ is reached by evaluating the
    series there.}
  \label{fig:omega-plane-contour}
\end{figure}

For DIS, it is natural to work in the variable $\o = 1/x$, in which the cut of
$T_a$ at fixed $Q^2<0$ lies at $|\o|\ge 1$, i.e.\
$\o\in[1,\infty)\cup(-\infty,-1]$.  The function is therefore analytic
in a neighbourhood of the origin $\o=0$ and admits a Taylor expansion
\be
  T_a(Q^2, \o) \;=\; \sum_{N=0}^{\infty} c_N(Q^2)\,\o^N,
  \qquad |\o|<1.
  \label{eq:omega_Taylor}
\ee
The Taylor coefficients are the $(N)$-th $\o$-derivatives of $T_a$
at the origin,
\be
  c_N(Q^2) \;=\; \frac{1}{(N)!}\,
  \partial_\o^{N}T_a(Q^2,\o)\Big|_{\o=0}.
  \label{eq:Taylor_coeff}
\ee
We now observe that these coefficients are expressed, without any
subtractions, by inserting the dispersive weight $x^N$ into the DIS
dispersion formula.  Using $\o=1/x$ and the discontinuity relation
$W_a = 2\,\mathrm{Im}[T_a]$,
\be
  c_N(Q^2) \;=\; \frac{1}{\pi}\int_0^1 dx\; x^{N-1}\,W_a(Q^2,x).
  \label{eq:cN_moment}
\ee
 The dispersion relation is depicted  in Fig.\ \ref{fig:omega-plane-contour} with the $\omega$ plane contours shown explicitly. 

The $N$-th $x_F$-Mellin transform of $T_a$ is obtained by integrating
the Taylor series~\eqref{eq:omega_Taylor} term by term:
\bea
  \int_0^1 dx_F\; x_F^{N-1}\, T_a(Q^2,x_F)
  &=& \sum_{k=0}^{\infty} c_k(Q^2)
  \int_0^1 dx_F\; x_F^{N+k-1}
  \;=\; \sum_{k=0}^{\infty} \frac{c_k(Q^2)}{N+k}.
  \label{eq:MellinT}
\eea
Here, the sum and integral can be interchanged because a Taylor series is uniformly convergent on a compact sub space of the domain of analyticity. 
Applying this $x_F$-Mellin transform to the LHS of Eq.\ (\ref{eq:final-fact-SIA})
\bea
  &&\frac{1}{|Q_2^2|-|Q_1^2|}
  \int_0^1 dx_F\;x_F^{N-1}
  \Bigl[T_a\!\left(-|Q_1^2|,x_F\right)-T_a\!\left(-|Q_2^2|,x_F\right)\Bigr]
  \nn \\
  &=&
  \frac{1}{|Q_2^2|-|Q_1^2|}
  \sum_{k=0}^{\infty}
  \frac{c_k\!\left(-|Q_1^2|\right)-c_k\!\left(-|Q_2^2|\right)}{N+k},
  \label{eq:LHS_Mellin}
\eea
a weighted sum of DIS Mellin moments~\eqref{eq:cN_moment} at the two
Euclidean scales.  Finally, using DIS factorization as in Eq.\ (\ref{eq:DIS_factorization}), we have

\bea
  c_k(Q^2,\mu_D^2)
  &=& \frac{1}{\pi}\int_0^1 dx\; x^{k-1}
      \sum_i\int_0^1\frac{d\xi}{\xi}\,
      C_{a,i}\!\left(\frac{x}{\xi},\frac{Q^2}{\mu_D^2}\right)
      f_i(\xi,\mu_D^2) \nn \\
  &=& \sum_i \widetilde{C}_{a,i}\!\left(k,\frac{Q^2}{\mu_D^2}\right)
      \tilde{f}_i(k,\mu_D^2),
  \label{eq:ck_factorized}
\eea
where we have defined the Mellin-projected DIS hard coefficient
\be
  \widetilde{C}_{a,i}\!\left(k,\frac{Q^2}{\mu_D^2}\right)
  \;\equiv\;
  \frac{1}{\pi}
  \int_0^1 dx\; x^{k-1}
  \int_0^1\frac{d\xi}{\xi}\,
  C_{a,i}\!\left(\frac{x}{\xi},\frac{Q^2}{\mu_D^2}\right),
  \label{eq:Ctilde_def}
\ee
and $\tilde{f}_i(k,\mu_D^2) = \int_0^1 d\xi\,\xi^{k-1}f_i(\xi,\mu_D^2)$
is the $k$-th Mellin moment of the PDF.
Substituting~\eqref{eq:ck_factorized} into~\eqref{eq:LHS_Mellin},
\bea
  &&\frac{1}{|Q_2^2|-|Q_1^2|}
  \sum_{k=0}^{\infty}
  \frac{c_k(-|Q_1^2|,\mu_D^2)-c_k(-|Q_2^2|,\mu_D^2)}{N+k}
  \nn \\
  &=&
  \sum_{k=0}^{\infty}\sum_i\,
  \mathcal{M}_{Nk}^{(a)}\!\left(Q_1^2,Q_2^2,\mu_D^2\right)
  \tilde{f}_i(k,\mu_D^2),
  \label{eq:LHS_factorized}
\eea
where we have defined the \emph{dispersive hard-coefficient matrix}
\be
  \mathcal{M}_{Nk}^{(a)}\!\left(Q_1^2,Q_2^2,\mu_D^2\right)
  \;\equiv\;
  \frac{\widetilde{C}_{a,i}\!\left(k,-|Q_1^2|/\mu_D^2\right)
       -\widetilde{C}_{a,i}\!\left(k,-|Q_2^2|/\mu_D^2\right)}
       {(|Q_2^2|-|Q_1^2|)(N+k)}.
  \label{eq:M_matrix}
\ee
Equating~\eqref{eq:LHS_factorized} with~\eqref{eq:final-fact-SIA} yields
the relation for the $N$-th Mellin moment of the fragmentation functions:
\bea
\sum_i
  \widehat{H}_{a,i}(N,Q_1^2,Q_2^2,\mu^2)\,
  \tilde{D}_i(N,\mu^2)\nn \\
  &&\hspace{-8cm}
  =\sum_{k=0}^{\infty}\sum_i\,
  \mathcal{M}_{Nk}^{(a)}\!\left(Q_1^2,Q_2^2,\mu_D^2\right)
  \tilde{f}_i(k,\mu_D^2)-\sum_i
  \widehat{K}_{a,i}(N,Q_1^2,Q_2^2,\mu^2).
  \label{eq:master_moment}
\eea
This is the second central result of this paper. The $N$-th Mellin moment of the fragmentation function $\tilde{D}_i(N,\mu_F^2)$
is determined by the full tower of PDF Mellin moments
$\{\tilde{f}_i(k,\mu_D^2)\}_{k\ge 0}$, with the matrix
$\mathcal{M}_{Nk}^{(a)}$ encoding the dispersively continued DIS hard
coefficients evaluated at the two Euclidean scales $Q_1^2$ and $Q_2^2$. Crucially, the only non-perturbative input on the right-hand side is the same set of parton distribution functions that enter the DIS factorization theorem — no new long-distance data is required.
The mixing between moment labels $k$ and $N$ is a direct consequence
of the dispersive kinematics: the $1/(N+k)$ kernel in~\eqref{eq:M_matrix}
arises from the term-by-term Mellin integration of the Taylor
series~\eqref{eq:omega_Taylor}, and is absent in any purely collinear
treatment.  In particular, $\mathcal{M}_{Nk}^{(a)}$ is not diagonal in
moment space. In addition, the fragmentation and the distribution function are related by the Mellin transformed residual function. So the Gribov-Lipatov-like identification $\tilde{D}_i(N)
\sim \tilde{f}_i(N)$ emerges only when the off-diagonal elements are
subleading and when the residual function is small. 
This concludes our treatment of cross-symmetric dispersion within the context of leading power factorization. 
\section{Disconnected cuts and factorization}
\label{sec:disconnected}
In this section, we discuss the origin of the ``residual term" that first appeared in Eq.\ (\ref{eq:crossing213}), and subsequently in dispersion relations that followed.  A factorization ansatz was made for this term in Eq.\ (\ref{eq:disc_factorization}), which rendered it calculable up to power corrections in perturbation theory. 
The derivation of the dispersion relation in Sec.~\ref{sec:results} rests on the
hadronic tensor being defined through the \emph{time-ordered} product of currents whose discontinuity in the $q^2>0$ region gave rise to two terms: semi inclusive annihilation and residual terms. Here, we first discuss cuts of the direct channel, then the crossed channel and the origin of the residual terms. A diagrammatic interpretation of the residual terms immediately leads to  leading power factorization of the residual function. 
\subsection{Analysis of cuts}
In the timelike (SIA) region: the time-ordered product
admits \emph{partially disconnected} cuts that are kinematically accessible for
$q^2>0$ and corrospond to incoming (anti) protons. Such cuts can be classified.

\paragraph{The four-point Green function and LSZ reduction.}
The fundamental object is the full time-ordered four-point Green function
\begin{equation}
\label{eq:G4}
    G^{(4)}(x_1, x_2, x_3, x_4)
    \;=\;
    \langle \Omega |\,
        \mathcal{T}\{\phi(x_1)\,\phi(x_2)\,J(x_3)\,J(x_4)\}
    \,|\Omega\rangle,
\end{equation}
where $\phi(x_1),\phi(x_2)$ are the proton field insertions and $J(x_3),J(x_4)$
are the current insertions. The forward Compton amplitude is obtained by LSZ
reduction on the two proton legs only,
\begin{equation}
\label{eq:LSZ-mom}
    \langle \vec{P},{\rm out}|\,\mathcal{T}\{J(x_3)J(x_4)\}\,|\vec{P},{\rm in}\rangle
    \;=\;
    \lim_{P^2 \to m_p^2}
    \frac{(P^2 - m_p^2)^2}{Z}
    \;\widetilde{G}^{(4)}(P,-P,x_3,x_4).
\end{equation}
The current insertions are not reduced. The factor $(P^2-m_p^2)^2/Z$ requires
$\widetilde{G}^{(4)}$ to supply a double pole in $P^2-m_p^2$ for the amplitude
to be nonzero.
\begin{figure}[ht]
    \centering
    \includegraphics[width=0.5\linewidth]{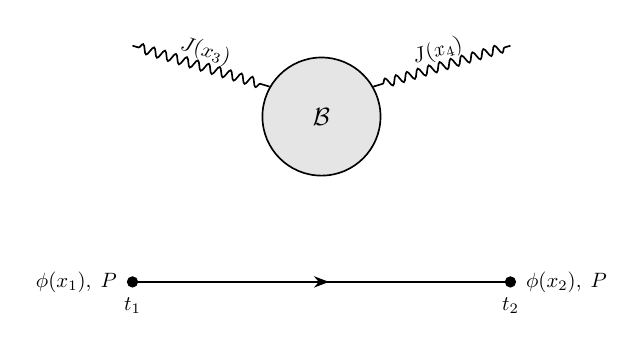}
    \caption{Fully disconnected graphs that potentially contribute to the imaginary part of $T^{\m\n}$. Cuts of this graph have singularities in $q^2$ alone, independently of the value of $x$. However, they don't survive LSZ reduction.  }
    \label{fig:fully-disconnected}
\end{figure}
We now observe that fully disconnected graphs like those in Fig.\ \ref{fig:fully-disconnected} only have a single pole in $P^2-m_p^2$ and as a result do not survive LSZ reduction. Therefore, cuts of this graph, which are not SIA processes and have singularities in $q^2$ alone, independent of the value of $x$, do not contribute to the imaginary part (or to the real part). 

\begin{figure}[ht]
    \centering
    \includegraphics[width=0.5\linewidth]{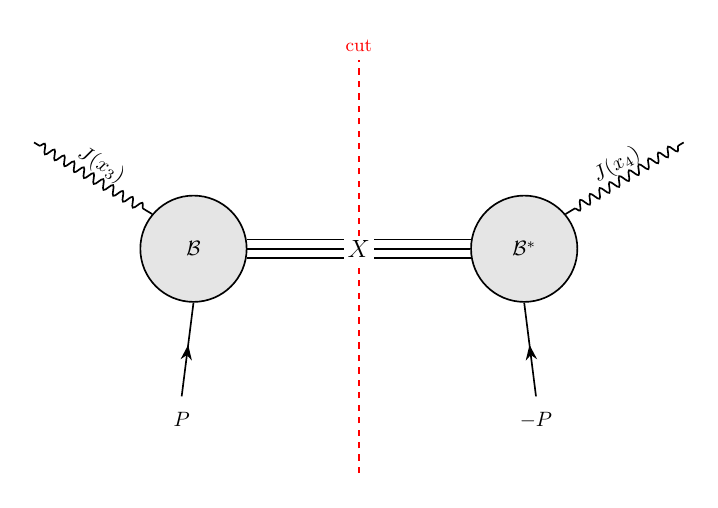}
    \caption{Graphs that are fully connected on either side of the cut. Such graphs, when crossed yield semi-inclusive annihilation cross-section. }
    \label{fig:fully-connected}
\end{figure}

\begin{figure}[h]
  \centering
  \begin{minipage}[c]{0.45\textwidth}
    \centering
    \includegraphics[width=\textwidth]{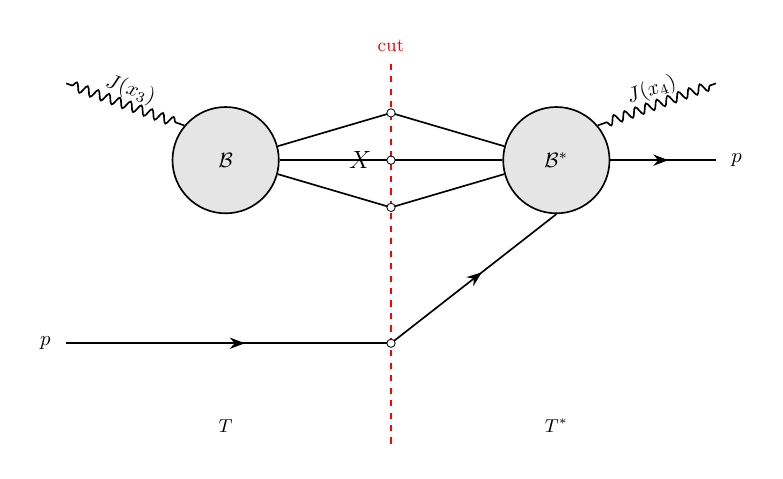}
    (a)
  \end{minipage}
  \hfill
  \begin{minipage}[c]{0.45\textwidth}
    \centering
    \includegraphics[width=\textwidth]{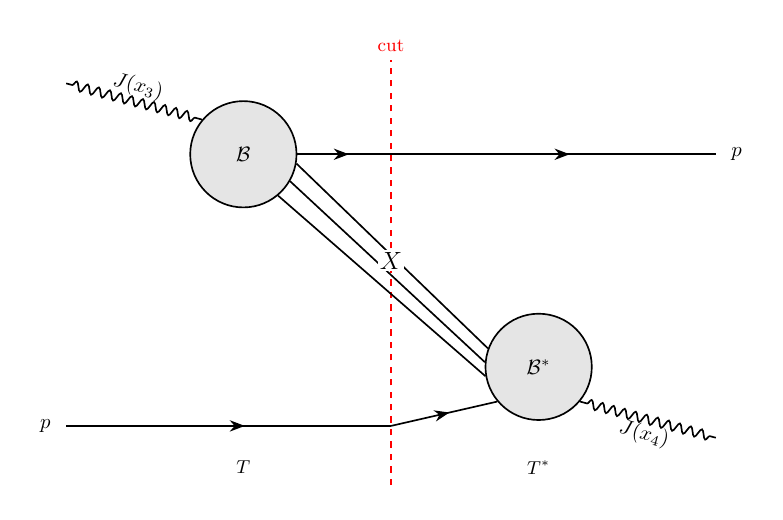}
    (b)
  \end{minipage}
  \caption{%
    A partially disconnected contribution to $\widetilde{G}^{(4)}$.
  }
  \label{fig:partly-connected}
\end{figure}

Let us now list the three remaining types of cuts of the time ordered product in the $q^2$. First are cuts that are connected on both side, as shown in Fig.\ \ref{fig:fully-connected}. These cuts, upon crossing will yield the SIA cross-section. 
 The next class of cuts  are shown in Fig.\ \ref{fig:partly-connected}. The goal of this section is to show that cuts of the sort shown in Fig.\ \ref{fig:partly-connected} which appear in the imaginary part of $T^{\m\nu
}$ combine to form the residual function. Such cuts, are extraneous to   the SIA cross section since the photon and proton are in separate disconnected graphs.  To build a diagrammatic interpretation of these cuts, we will study crossing in time-ordered-perturbation-theory. 
In TOPT, $\widetilde{G}^{(4)}$ is represented as a sum over time-orderings of
interaction vertices, $v$, with each line carrying a definite on-shell energy
$\eta_{l,v}\,\omega_{\vec{q}_l}$, $\eta_l=\pm 1$, where $l$ is a line emanating to or from $v$ ~\cite{Sterman:2023xdj}. Here, $\vec{q}_l$ is to be understood as the line momentum of the line $l$, which is a linear combination of the loop three-momenta $\vec{k}_i$. The
schematic form is
\begin{equation}
\label{eq:TOPT-G4}
    \widetilde{G}^{(4)}
    \;=\;
    \int \prod_{\text{loops}} \frac{d^3k_i}{(2\pi)^3}
    \prod_{\text{lines}\;l} \frac{1}{2\omega_{\vec{q}_l}}
    \;\sum_{\text{orderings}}
    \prod_{\text{vertices}\;v}
    \int dt_v\;
    e^{-it_v\, \mathcal{E}_v}
    \;\mathcal{N},
\end{equation}
where $\mathcal{N}$ collects coupling and numerator factors and the sum over orderings, is to be understood as a sum over two possible orientations for each line.-- one forward moving and the other moving back in time.  
\begin{equation}
\label{eq:Ev}
    \mathcal{E}_v
    \;=\;
    \sum_{\alpha\in\{P,Q\}}\bar{\eta}_{\alpha,v}\, \alpha^0
    \;+\;
    \sum_{l \ni v} \eta_{l,v}\,\omega_{\vec{k}_l},
\end{equation}
with $\bar{\eta}_{\alpha,v}\in\{-1,0,+1\}$ recording whether $\alpha^0$ flows
into or out of vertex $v$. Two vertices in~\eqref{eq:TOPT-G4} are special:
$t_1$ and $t_2$, at which the reduced proton lines attach. 
the truncated proton line itself — so $\mathcal{E}_{t_1} = P^0 - \omega_{\vec{P}}$
exactly. The $t_1$ integral is therefore
$
    \int dt_1\; e^{-it_1(P^0 - \omega_{\vec{P}})}.$

When $t_1$ is first in the time ordering the lower limit is $-\infty$, and the
integral evaluates to
\begin{equation}
\label{eq:t1-pole}
    \int_{-\infty}^{t_a} dt_1\; e^{-it_1(P^0 - \omega_{\vec{P}})}
    \;=\;
    \frac{i\,e^{it_a(P^0-\omega_{\vec{P}})}}{P^0 - \omega_{\vec{P}} + i\varepsilon},
\end{equation}
where $t_a$ is the next vertex in the ordering and the residual phase is absorbed
there. The pole $(P^0-\omega_{\vec{P}})^{-1}$, 
combines with the LSZ reduction factor $Z^{-\frac{1}{2}}((P^{0})-\omega_{\vec{P}})2\omega_{\vec{P}}$ to leave  $Z^{-\frac{1}{2}}2\omega_{\vec{P}}$ in the on-shell $P^0\rightarrow\omega_{\vec{P}}$ limit ~\eqref{eq:LSZ-mom}. Finally, this combines with $\frac{Z^{1/2}}{2\omega_P}$ coming from self energy corrections in the graph, to truncate the external leg as in covariant perturbation theory. However, one key distinction from covariant perturbation theory is the requirement that the retained time orders necessarily admit a representation where the $t_1$ integral is unbounded from below. By the same argument $t_2$ integral must by unbounded from above to survive the LSZ reduction. We are now ready to discuss the status of the amplitude on cuts of the type shown in Fig.\ \ref{fig:partly-connected} which disconnect the either the amplitude or the complex conjugate amplitude in the direct channel. Using the standard perturbative unitarity arguments \cite{Sterman:1993hfp,Sterman:2023xdj} we may write
\begin{multline}
  \text{Im}\!\left(\langle\vec{P},{\rm out}|\,
    \mathcal{T}\{J(x_3)J(x_4)\}\,
  |\vec{P},{\rm in}\rangle\right)
  = \sum_{C}\sum_{O'_L,O'_R}
    \mathcal{N}_{O'_L\cup O'_R}
    \int \prod_{\text{loops}} \frac{d^3k_i}{(2\pi)^3}
    \prod_{\text{lines}\;l} \frac{1}{2\omega_{\vec{q}_l}}
  \\
  \times\; a_{L}(\vec{k}_{i,L},Q,P)\;
    2\pi\delta\!\left(Q^0+P^0-\sum_{i\in C} \omega_i\right)\;
    a_R(\vec{k}_{i,L},Q,P)\;\frac{(P^2-m_P^2)^2}{Z},
\end{multline}
where $a_{L,R}$ are the amplitudes on the left and the right of the cut $C$  with the numerator factors stripped and collected in $\mathcal{N}_{O'_L\cup O'R}$. The orders $O'_L,O'_R$ are the subset of all orders in the graph with  restriction that the time $t_1$ is unbounded from below on the left and the $t_2$ is unbounded from above on the right.  The cut sum has been restricted to cuts that occur after $Q^0$ has entered the  graph, under the assumption that the proton is the lightest stable baryon and cuts of the form $\delta \left(P^0-\sum_{i\in C} \omega_i\right)$ have no phase space support (admit vanishing amplitudes since $\sum_i \omega_i>m_p$). As a result, time orders in which  the vertex where $-Q$ enters the graph prior to $Q$ also have no phase space support because their energy conserving functions are of the form $\delta\left(-Q^0+P^0-\sum\limits_{i\in C} \omega_i\right)$.

Consider for example, cuts of the type shown in the left panel of Fig.\ \ref{fig:partly-connected} which disconnects the amplitude but not the complex conjugate amplitude. It was shown in \cite{Sterman:2023xdj} that when $a_L$ consists of two or more disconnected components, $a_L$ may be written as
\begin{multline}
  (P^0-\o_P)\,a_{L}(\vec{k}_{i,L},Q,P)\;
    2\pi\delta\!\left(Q^0+P^0-\sum_{i\in C}\omega_i\right)
  \\
  = \prod_{C_q\in P_1}\frac{i(P^0-\o_P)}{Q^0-\sum_{i\in C_q}\o_i+i\e}
    \times \prod_{C_p\in P_2}
      \frac{i}{P^0-\sum_{i\in C_p}\o_i+i\e}
    \times \frac{i}{P^0-\omega_{\vec{P}}+i\e}\\
    \times \frac{i}{Q^0-\sum_{i\in C\cup C_q}\omega_i+i\e}
  \times\; 2\pi\delta\!\left(Q^0+P^0-\sum_{i\in C}\omega_i\right)
    \left(Q^0+P^0-\sum_{i\in C}\omega_i\right),
  \label{eq:disconnected-graphs}
\end{multline}
where $P_1$ is the connected subgraph containing $q$, $P_2$ is the subgraph containing $P$ and cuts of $P_2$ are self energy corrections of the proton. The last line of Eq.\ \ref{eq:disconnected-graphs} is clearly the distribution with no support almost everywhere (except in measure zero corners of phase space) since it is of the form $x\delta(x)$. The only way for the resulting distribution to  have support on a set of finite measure is if both denominators $(P^0-\omega_{\vec{P}}+i\e)^{-1}$ and $(Q^0-\sum\limits_{i\in C\cup C_q}\omega_i+i\e)$ both vanish simultaneously, i.e. momentum is conserved in both disconnected components independently.  Therefore, in the direct channel, prior to crossing, disconnected cuts do not vanish for $q^2>0$. When $q^2<0$, such cuts do vanish because of the spectral property-- the energy of the final state $X$ is necessarily positive. 

Let us now discuss the crossed channel, where $\tilde{G}^{(4)}(q,\bar{P}=-P)$ is analytically continued to a region where the $\bar{P^0}$ is positive. In this region, the set of time orders of the correlation function that survive LSZ reduction have poles of the form $(\bar{P^0}-\omega_{\bar{P}})2\omega_{\vec{P}}=(-{P^0}-\omega_{{P}})2\omega_{\vec{P}}$. This is a complimentary set of time orders. However, such time orders are necessarily in one-to-one correspondence with time orders from the direct channel. In this case, the $t_1$ integral when unbounded from above reads:
\bea
 \int^{\infty}_{t_b} dt_1\; e^{-it_1(P^0 - \omega_{\vec{P}})}
    \;=\;
    \frac{i\,e^{it_b(P^0-\omega_{\vec{P}})}}{-P^0 - \omega_{\vec{P}} + i\varepsilon},
\eea
which is the pole needed to cancel LSZ reduction zero. As a result, time orders of the correlation function that satisfy reduction are those with $t_1$ being unbounded from above. Furthermore, every time order in which the  $t_1$ integral is unbounded from above and the $t_2$ from below, the time order contributes to the LSZ reduced amplitude. Therefore time orders in the crossed channel are in one-to-one correspondence with the time orders in the direct channel. If $(t_1,\pi_1(\{t_{a_i}\}), \pi_2(\{t_{a_i}\})\ldots , t_2)$ is a time order in the direct channel, $(t_2,\pi_1(\{t_{a_i}\}), \pi_2(\{t_{a_i}\})\ldots , t_1)$ represents a time order that survives LSZ reduction in the crossed channel. 

Every time order, and as a result, every cut in the crossed channel, can be generated from the direct channel by transposing the incoming leg $P$ which was an ``early" vertex to a ``late" vertex which appears as a final state while simultaneously moving the vertex where $P$ is outgoing to the in-state. We call this move in every time order, a crossing move.  Having already classified cuts in the direct channel, it is simple to organize cuts in the crossed channel. 

Let us now move on to analyzing the crossed channel. We divide cuts into three categories:
\begin{itemize}
    \item Cuts of type (i), which are cuts shown in Fig. \ref{fig:fully-connected}, in the crossed channel.
    \item Cuts of type (ii), which are cuts shown in Fig. \ref{fig:partly-connected} (a), in the crossed channel.
    \item Cuts of type (iii), which are cuts shown in Fig. \ref{fig:partly-connected} (b), in the crossed channel.
\end{itemize}
We have therefore partitioned the imaginary part of the graph in the crossed channel into three pieces, following the application of perturbative unitarity. For brevity we write
\be
  \ImCorr{\bar{P}} \;\equiv\;
  \text{Im}\!\left(\langle{-\vec{\bar{P}},{\rm out}}|\,
    \mathcal{T}\{J(x_3)J(x_4)\}\,
  |{-\vec{\bar{P}},{\rm in}}\rangle\right),
\ee
so that the partition reads
\begin{align}
  \ImCorr{\bar{P}}
    &= \sum_{C^{(i)}}  \ImCorr{\bar{P}}\Big|_{(i)}
     + \sum_{C^{(ii)}} \ImCorr{\bar{P}}\Big|_{(ii)}
     + \sum_{C^{(iii)}}\ImCorr{\bar{P}}\Big|_{(iii)},
\end{align}
where $|_{(x)}$ denotes restriction to cuts of type $x$ in the crossed channel. 
Let us now analyze cuts of type (i) which may be written as:
\begin{multline}
  \sum_{C^{(i)}}\ImCorr{\bar{P}}\Big|_{(i)}
  = \sum_{C^{(i)}}\sum_{O'_L,O'_R}
    \mathcal{N}_{O'_L\cup O'_R}
    \int \prod_{\text{loops}} \frac{d^3k_i}{(2\pi)^3}
    \prod_{\text{lines}\;l} \frac{1}{2\omega_{\vec{q}_l}}
  \\
  \times\; a_{L}(\vec{k}_{i,L},Q,-\bar{P})\;
    2\pi\delta\!\left(Q^0-\bar{P}^0-\sum_{i\in C} \omega_i\right)\;
    a_R(\vec{k}_{i,L},Q,-\bar{P})\;\frac{(P^2-m_P^2)^2}{Z},
\end{multline}
where $a_{L,R}$ are connected graphs. These graphs correspond to SIA and yield $\bar{W}_a$ term in Eq.\ (\ref{eq:crossing213}).

Let us now argue that partly-disconnected cuts of the type in Fig.\ \ref{fig:partly-connected} yield non-vanishing contributions after crossing. Cuts of type $(ii)$, upon crossing yield
\begin{multline}
  \sum_{C^{(ii)}}\ImCorr{\bar{P}}\Big|_{(ii)}
  = \sum_{C^{(ii)}}\sum_{O'_L,O'_R}
    \mathcal{N}_{O'_L\cup O'_R}
    \int \prod_{\text{loops}} \frac{d^3k_i}{(2\pi)^3}
    \prod_{\text{lines}\;l} \frac{1}{2\omega_{\vec{q}_l}}
  \\
  \times\; a_{L}(\vec{k}_{i,L},Q)\;
    2\pi\delta\!\left(Q^0-\sum_{i\in C} \omega_i\right)\;
    a_R(\vec{k}_{i,L},Q,-\bar{P})\;\frac{(P^2-m_P^2)^2}{Z}
    \;+\;\text{c.c.}
\end{multline}
In this expression, we have utilized the argument for independent momentum conservation on connected components to write
\bea
2\pi\delta\left(Q^0+\bar{{P}}^0-\omega_{\bar{P}}-\sum\limits_{i\in C} \omega_i\right)=2\pi\delta\left(Q^0-\sum\limits_{i\in C} \omega_i\right).
\eea
Finally, on cuts of type $(iii)$, the imaginary part reads: 
\begin{multline}
  \sum_{C^{(iii)}}\ImCorr{\bar{P}}\Big|_{(iii)}
  = \sum_{C^{(iii)}}\sum_{O'_L,O'_R}
    \mathcal{N}_{O'_L\cup O'_R}
    \int \prod_{\text{loops}} \frac{d^3k_i}{(2\pi)^3}
    \prod_{\text{lines}\;l} \frac{1}{2\omega_{\vec{q}_l}}
  \\
  \times\; a_{L}(\vec{k}_{i,L},Q,\bar{P})\;
    2\pi\delta\!\left(Q^0+\bar{P}^0-\sum_{i\in C} \omega_i\right)\;
    a_R(\vec{k}_{i,L},Q,-\bar{P})\;\frac{(P^2-m_P^2)^2}{Z}.
\end{multline}
All of these contributions are non-vanishing and cuts of type $(ii),(iii)$ are extraneous to the SIA cross-section. We therefore recognize that
\begin{align}
  \Gamma(q^2,q\cdot \bar{P})
    &= \sum_{C^{(ii)}} \ImCorr{\bar{P}}\Big|_{(ii)}
     + \sum_{C^{(iii)}}\ImCorr{\bar{P}}\Big|_{(iii)}.
\end{align}
This concludes the diagrammatic cut analysis of $T^{\mu\n}$ in the $q^2\geq 0$ region. In the next subsection, we study the factorization properties of $\Gamma(q^2,q\cdot \bar{P})$.

\subsection{Factorization of the residual function}
\label{subsec:fact-residual}
\begin{figure}[h]
  \centering
  \includegraphics[width=\textwidth]{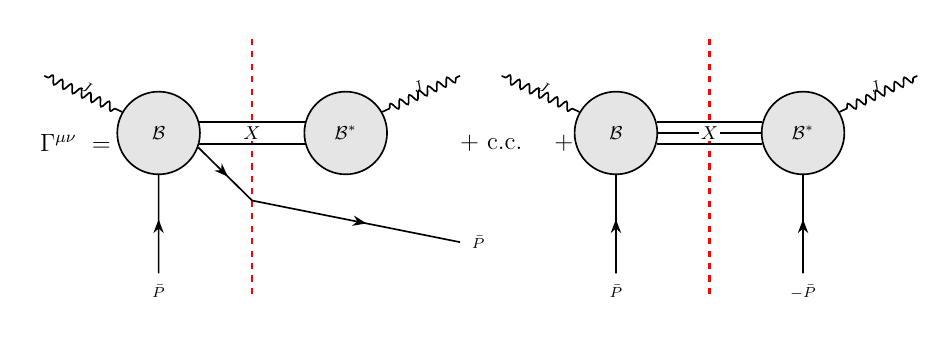}
  \caption{Diagrammatic representation of the residual function
    $\Gamma^{\mu\nu}(q^2, q\cdot\bar{P})$ as a sum of two classes of
    partially disconnected cuts in the crossed channel.
    \textit{Left:} A type~(ii) cut in which $\bar{P}$ enters and exits
    the amplitude-side blob $\mathcal{B}$, passing through the cut as a
    free line, while the complex-conjugate blob $\mathcal{B}^*$ couples
    only to the current $J$ and the final state $X$. The conjugate
    contribution (c.c.) is included.
    \textit{Right:} A type~(iii) cut in which $\bar{P}$ enters both
    $\mathcal{B}$ and $\mathcal{B}^*$ symmetrically from below, with
    each blob also coupled to its respective current insertion $J$.
    In both cases the final state $X$ connects the two blobs across the
    cut. These contributions are extraneous to the SIA cross section and
    together constitute the residual function $\Gamma^{\mu\nu}$.}
  \label{fig:gamma-eq}
\end{figure}

We begin our analysis of the residual function, having identified the diagrammatic contributions to it. The diagrams that contribute to timelike DIS structure functions which admit a tensor decomposition into two scalar functions as before
\bea
\Gamma^{\m\n}(q^2,q\cdot \bar{P})&=&\sum_{C^{(ii)}\cup C^{(iii)}}\text{Im}\left(\langle -\vec{\bar{P}},{\rm out}|\,\mathcal{T}\{J^\m(x_3)J^\n(x_4)\}\,|-\vec{\bar{P}},{\rm in}\rangle\right)\Bigg|_{(ii),(iii)}\label{eq:gamma-eq},
\eea
The diagrammatic version of this equation is shown in Fig.\ref{fig:gamma-eq}. 
 The diagrams
contributing to $\Gamma^{\mu\nu}$ are partially disconnected cuts of
the type~(ii) and type~(iii) shown in Fig.~\ref{fig:partly-connected};
they admit the same tensor decomposition as the hadronic tensor,
\bea
\Gamma^{\m\n}(q^2,q\cdot \bar{P})
  &=& \left(\frac{q^{\m}q^{\n}}{q^2}-\eta^{\m\n}\right)
       \Gamma_{1}(q^2,2\bar{P}\cdot q)
  \nn \\ &&
  -\frac{x_F^2}{4}
       \left(\frac{2}{x_F}\bar{P}^\m-q^{\m}\right)
       \left(\frac{2}{x_F}\bar{P}^\n-q^{\n}\right)
       \frac{\Gamma_2 (q^2,2 \bar{P}\cdot q)}{\bar{P}\cdot q},
  \label{eq:gamma-tensor}
\eea
so that $\Gamma_a$ ($a=1,2$) are the two scalar residual functions.
To demonstrate that the residual function $\Gamma_a$ factorizes, we evaluate the matrix element in Eq.\ (\ref{eq:gamma-eq}) with incoming (and outgoing) partons, classify the most general pinch surface, write an expression near each pinch, hold the jet cut fixed and sum over all soft subdiagram cuts and find that the $\bar{P}$ jet facotrizes into the familiar DIS collinear jet function and a hard function  which factorizes.

\paragraph{Leading-region decomposition.}
Working in light-cone coordinates $(+,-,\perp)$, with $\bar{P}^{\m}=(P^+,0^-,0_{\perp})$,
we identify the leading regions $L$ in loop-momentum space that
contribute to $\G^{\m\n}$ at leading power in $Q^2$.  We first decompose $\Gamma$ into two pieces, one piece due to cuts of type (iii) and the second piece from cuts of type (ii),
\bea
\Gamma^{\m\nu}=\Gamma^{\m\nu,(1)}+\Gamma^{\m\nu,(2)},\label{eq:gamma-decomp}
\eea
such that each term is defined by the topology of the cut
\begin{align}
\Gamma^{\m\nu,(1)}&=\frac{1}{8\pi}&\sum_X \mathcal{M}\left(j^{\m}(q)+\bar{P} \rightarrow X\right)\mathcal{M}^*\left(j^{\n}(-q)-\bar{P} \rightarrow X\right) (2\pi)^4\d^4(q+\bar{P}-P_x), \label{eq:gamma-algorithm}\\
\Gamma^{\m\nu,(2)}&=\frac{1}{8\pi}&\sum_X \mathcal{M}\left(j^{\m}(q)+\bar{P} \rightarrow X+\bar{P}\right)\mathcal{M}^*\left(j^{\n}(-q) \rightarrow X\right)(2\pi)^4\d^4(q-P_x)+\text{c.c}.\nn
\end{align}
Beginning the analysis of $\Gamma^{\m\n, (1)}$, we study the partonic cross section  in the lightcone limit. As usual, the leading regions of the loop momenta organize into four gauge-invariant subgraphs.
The \emph{hard subgraphs} $H_{\r\s}$ on the left and the right of the cut contain the two current insertions
$J^\r$ and $J^\s$ resppectively, together with all internal lines whose virtuality is
of order $Q^2$; it is never cut, since no real cut can put an off-shell
line of virtuality $\sim Q^2$ on shell while conserving the external
momenta.  The \emph{$\bar{P}$-jet subgraph} $J_{\bar{P}}$ contains all
lines collinear to $\bar{P}$, carrying large plus momenta $\sim Q$ and
small minus and transverse momenta.  The \emph{soft subgraph} $S$
contains lines whose all momentum components are small, of order
$\l \ll Q$.  Finally, there are $N_j$ \emph{final-state jet subgraphs}
$\{J_j\}$, one for each collinear direction of the final-state
particles in $X$, each carrying large plus momenta of order $Q$. A cut is a choice of cut lines in the $\bar{P}$ jet, the soft subgraph and the final state jet subgraph.

The soft subgraph connects to $J_{\bar{P}}$ by exchanging gluons
carrying Lorentz indices $\{\m_l\}$ and momenta $\{q_l\}$, and
connects to the final-state jets by exchanging gluons carrying indices
$\{\n_l\}$ and the same momenta $\{q_l\}$ (momentum is conserved at
each soft vertex).  The hard subgraph receives a single collinear line
from $J_{\bar{P}}$ on either side of the cut, carrying plus momentum $k_{\bar{P}}^+\sim Q$, and a
single collinear line on either side of the cut from each final-state jet $J_j$ .  The contribution from region $L$ of a graph $G$ to the
residual tensor, summed over cuts consistent with the type (iii)
restriction of Sec.~\ref{sec:disconnected}, is
\bea
  \G^{\r\s,(1)}
  &=&
  \sum_{G,G_L}\sum_{C_f}\sum_{C_S}\sum_{C_{\bar{P}}}
  \int \frac{dk_{\bar{P}}^+}{2\pi}
  \prod_{j=1}^{N_j}\int\frac{d(k_j\cdot \bar{n}_j)}{2\pi}
  \prod_l \int \frac{d^4 q_l}{(2\pi)^4}
  \nn \\ &&\hspace{0.5cm}
  \times\;
  \left[J_{\bar{P}}^{(C_{\bar{P}})}\right]^{\m_1\cdots\m_n}\!\!\left(k_{\bar{P}}^+, \{q_l\}\right)
  \;S_{\m_1\cdots\m_n}^{{(S_C)}\quad\;\n_1\cdots\n_m}\!\left(\{q_l\}\right)
  \nn \\ &&\hspace{0.5cm}
  \times\;
  \prod_{j=1}^{N_j}
  \left[J_j\right]^{(C_f)}_{\n_{l_j}\cdots}\!\!\left(k_j\cdot \bar{n}_j, \{q_l\}\right)
  \;H^{\,\r\s}\!\left(k_{\bar{P}}^+, \{k_j\}\right),
  \label{eq:Gamma1_region}
\eea
where $n+m$ is the number of soft gluons, the Lorentz indices
$\m_1\cdots\m_n$ on $J_{\bar{P}}$ and $S$ are contracted, and
$\n_1\cdots\n_m$ on $S$ and $\{J_j\}$ are contracted.  The
integrations over $k_{\bar{P}}^+$ and $\{k_j\cdot \bar{n}_j\}$ represent the
single collinear line connecting each jet to the hard subgraph.  The
soft momenta $\{q_l\}$ are integrated over the soft region
$q_l^+, q_l^-, |\mathbf{q}_l| \ll Q$. Therefore, $\bar{n}_j$ is to be interpreted as the unit vector in the opposite lightcone direction of each jet.  To obtain the scalar residual functions
$\G_a$, one projects with $\mathcal{P}_{a,\r\s}$, the standard
projector on to the $a$-th tensor structure of Eq.~\eqref{eq:gamma-tensor}. Here, the hard functions on either side of the cut have been combined. 

\paragraph{Grammer--Yennie decomposition.}
Each final-state jet subgraph $J_j$ receives soft gluon attachments carrying
momenta $\{q_l\}$ from the soft subgraph $S$.  At leading power, the
virtuality of lines inside $J_j$ is of order $k_j^2 \sim \lambda^2 Q^2 \ll
Q^2$, so each soft gluon with $q_l^2 \sim \lambda^4 Q^2$ attaches to a
nearly on-shell collinear line.  Applying the Grammer--Yennie
decomposition~\cite{Grammer:1973db} to each such attachment replaces
\begin{equation}
  \label{eq:GY}
  g^{\mu\nu}
  \;\longrightarrow\;
  K^{\mu\nu}(q_l,n_j) + G^{\mu\nu}(q_l,n_j),
\end{equation}
where the $K$-part is leading power and the $G$-part is power suppressed and
dropped.  The $K$-tensor projects the soft gluon polarization onto the
collinear direction $n_j$,
\begin{equation}
  \label{eq:K_projector}
  K^{\mu\nu}(q_l,n_j)
  \;=\;
  \frac{q_l^\mu\, n_j^\nu}{q_l \cdot n_j},
\end{equation}
so that each retained soft gluon couples to $J_j$ through the eikonal vertex
$g_s\, n_j^\nu / (n_j\cdot q_l + i\varepsilon)$.

\paragraph{Summing over soft gluon attachments to the final-state jets.}
With the eikonal vertices identified, we sum over all ways of attaching the
fixed set of soft gluons $\{q_l\}$ to the collinear lines of $J_j$, holding
the soft momenta fixed.  For a single soft gluon $q_l$, the sum over
insertion points along a collinear fermion line is a telescoping sum of
eikonal propagators that collapses by partial fractions,
\begin{equation}
  \label{eq:eikonal_sum}
  \sum_{\text{insertions on }J_j}
  [J_j(q_l)]
  \;=\;
  \frac{g_s\, n_j \cdot \varepsilon(q_l)}{n_j\cdot q_l + i\varepsilon}
  \times [J_j],
\end{equation}
leaving the jet amplitude $[J_j]$ multiplied by a single overall eikonal
factor~\cite{Sterman:1993hfp}.  This is the eikonal identity;
it holds diagram by diagram and is a purely algebraic consequence of the
collinear kinematics and the $K$-projection.  For the full set of $n$ soft
gluons the identity iterates, since each gluon sees the same collinear
denominator structure,
\bea
  \label{eq:jet_eikonal_n}
  \sum_{G_L,G_R}
  [J_j(q_1,\ldots,q_n)]
  \;&=&\;
  \left[\prod_{l=1}^{n_l}
  \frac{g_s\, n_j\cdot A(q_l)}{n_j\cdot q_l + i\varepsilon}\right]
  \times [J_j] \left[\prod_{l=1}^{n_r}
  \frac{g_s\, n_j\cdot A(q_l)}{n_j\cdot q_l - i\varepsilon}\right] \\
  \;&=& \;
  \mathcal{W}_{n_j}[\{q_l\}]\times [J_j] \times  \mathcal{W}^{\dagger}_{n_j}\bigl[\{q_l\}\bigr],\nn 
\eea
where $\mathcal{W}_{n_j}[\{q_l\}]$ denotes the path-ordered product of
eikonal vertices for the fixed set $\{q_l\}$ and the same manipulation on the complex conjugate amplitude yield the conjugate Wilson line.  The Wilson line operator
\begin{equation}
  \label{eq:jet_WL}
  \mathcal{W}_{n_j}
  \;=\;
  \mathbf{P}\exp\!\left(
    ig_s \int_0^\infty d\lambda\; n_j \cdot A(\lambda n_j)
  \right)
\end{equation}
is simply the compact notation for this path-ordered product.  The jet $J_j$ has therefore completely factored from its soft attachments,
and the result of the sum is
\begin{equation}
  \label{eq:jet_after_soft_sum}
  \sum_{\text{attach.\ to }J_j}
  [J_j]^{(C_f)}_{\nu_{l_j}\cdots}
  \;=\;
  \mathcal{W}_{n_j}[\{q_l\}]_{\n_{lj}\ldots}
  \;\times\;
  [J_j(k_j^+)]^{(C_f)} \mathcal{W}^{\dagger}_{n_j}[\{q_l\}]_{\n_{lj}\ldots},
\end{equation}
where $J_j(k_j^+)$ depends only on the total plus-momentum it sources to the
hard part, and $S$ collects the remaining soft propagators and color matrices
connecting to the $\bar{P}$-jet.

\paragraph{Cancellation of the soft function.}
The soft function $S^{\nu_1\cdots\nu_m}_{\mu_1\cdots\mu_n}$ connects to the
$\bar{P}$-jet via indices $\mu_1\cdots\mu_n$ and to the final-state jets via
$\nu_1\cdots\nu_m$.  Upon summing over all cuts of $S$  --- i.e.\ inclusively over the soft final state ---
virtual and real soft emissions cancel by the KLN
theorem~\cite{Kinoshita:1962ur,Lee:1964is}, so that
\bea
  \label{eq:soft_unity}
  \sum_{C_S}
  \prod_{j} \mathcal{W}_{n_j}[\{q_l\}]_{\n_{lj}\ldots}S^{\nu_1\cdots\nu_m}_{\mu_1\cdots\mu_n}\mathcal{W}^{\dagger}_{n_j}[\{q_l\}]_{\n_{lj}\ldots}(\{q_l\})\left[J_{\bar{P}}^{(C_{\bar{P}})}\right]^{\m_1\cdots\m_n}\!\!\left(k_{\bar{P}}^+, \{q_l\}\right)\nn\\
  \;=\;
  \mathbf{1}\left[J_{\bar{P}}^{(C_{\bar{P}})}\right]\!\!\left(k_{\bar{P}}^+, \right),
\eea
leaving no residual soft factor at leading power. Here the right hand side of this equation is the identity in color space. 

\paragraph{The residual $\bar{P}$-collinear jet.}
After setting the soft function to unity, the contribution to
$\Gamma^{\rho\sigma,(1)}$ from region $L$ reduces to
\begin{align}
  \label{eq:Gamma1_after_soft}
  \Gamma^{\rho\sigma,(1)}\big|_L
  \;&=&\;
 \sum_{C_{\bar{P}}C_f} \int \frac{dk^+_{\bar{P}}}{2\pi}\,
  \Bigl[
    J^{(C_{\bar{P}})}_{\bar{P}}\!\left(k^+_{\bar{P}}\right)
  \Bigr]
  \;\times\;
  \prod_{j=1}^{N_j}\int \frac{d(k_j\cdot\bar{n}_j)}{2\pi}\,
  \Bigl[J_j^{(C_f)}(k_j\cdot\bar{n}_j)\Bigr]\nn \\ &&
  \;\times\;
  H^{\rho\sigma}\!\left(k^+_{\bar{P}},\{k_j\}\right),
\end{align}
where $H^{\rho\sigma}$ is the hard subgraph with the power-suppressed
$G$-parts dropped.

\paragraph{Identification of the collinear jet with the DIS PDF.}
The $\bar{P}$-jet subgraph $J_{\bar{P}}$ consists of a collinear line of
plus-momentum $k^+_{\bar{P}} = \xi P^+$ emerging from $|\bar{P}\rangle$.
Its discontinuity summed over cuts is
\begin{equation}
  \label{eq:jet_is_PDF}
  \sum_{C_{\bar{P}}}
  \bigl[J^{(C_{\bar{P}})}_{\bar{P}}\bigr](\xi P^+)
  \;=\;
 \int \frac{d\l}{2\pi}e^{-i\l \xi \bar{P}\cdot n} \langle \bar{P} |\,
    \bar\psi(\l n)\,{\gamma^+}\,
    \mathcal{W}_{n^-}(\l n,0)\psi(0)
  \,|\bar{P}\rangle
  \;\equiv\; f_i(\xi,\mu_r^2).
\end{equation}
The operator on the right-hand side is identical to the one appearing in the
DIS factorization theorem~\eqref{eq:DIS_factorization}: the same projector
$\gamma^+$ and the same hadronic state continued to the crossed channel.
The identification is made at the factorized level by identifying every graph that appears on the right hand side also appears on the left hand side and vice-versa. 

\paragraph{Assembled factorized expression.}
Substituting~\eqref{eq:jet_is_PDF} into~\eqref{eq:Gamma1_after_soft} and
summing over parton species $i$ and all leading regions gives
\begin{equation}
  \label{eq:Gamma1_factorized}
  \Gamma^{\rho\sigma,(1)}\left(q^2,x_F=\frac{1}{\o_F}\right)
  \;=\;
  \sum_i
  \int_0^1 \frac{d\xi}{\xi}\;
  \hat{K}^{\rho\sigma,(1)}_{i}\!\left(\frac{\o_F}{\xi},\frac{q^2}{\mu_r^2}\right)
  f_i(\xi,\mu_r^2)
  \;+\; \mathcal{O}\!\left(\frac{1}{Q^2}\right),
\end{equation}
where
\begin{equation}
  \label{eq:K_coeff}
  \hat{K}^{\rho\sigma,(1)}_{i}\!\left(z,\frac{q^2}{\mu_r^2}\right)
  \;=\;
  \prod_{j=1}^{N_j}\int \frac{d(k_j\cdot\bar{n}_j)}{2\pi}\,
  [J_j^{(C_f)}(k_j\cdot\bar{n}_j)]
  \;\times\;
  H^{\rho\sigma}\!\left(z P^+,\{k_j\}\right),
\end{equation}
is computable order by order in $\alpha_s$.  Projecting onto the scalar
structures $\mathcal{P}_{a,\rho\sigma}$ of Eq.~\eqref{eq:gamma-tensor}
yields the scalar coefficients $K_{a,i}(z,q^2/\mu_r^2)$ appearing in the
factorization ansatz~\eqref{eq:disc_factorization}, establishing
Eq.~\eqref{eq:disc_factorization} for $\Gamma_a^{(1)}$.

\paragraph{IR finiteness of the connected contribution to $\Gamma^{\mu\nu,(2)}$.}
We first show that the connected graphs contributing to $\Gamma^{\mu\nu,(2)}$ are IR
finite. The leading-region analysis proceeds as in Sec.~\ref{subsec:fact-residual},
with the same four gauge-invariant subgraphs: a $\bar{P}$-jet $J_{\bar{P}}$, final-state
jets $\{J_j\}$, a soft subgraph $S$, and a hard subgraph $H^{\rho\sigma}$. The key
distinction is that in $\Gamma^{\mu\nu,(2)}$, the $\bar{P}$-jet has $\bar{P}$ entering
\emph{and} exiting in the amplitude; it therefore cannot exchange a finite-energy collinear parton with
the hard subgraph without violating momentum conservation. Near the leading pinch,
$J_{\bar{P}}$ connects to the rest of the graph only through soft exchanges, and the
contribution from region $L$ takes the form
\begin{align}
    \Gamma^{\rho\sigma,(2)}\big|_L
    &=
    \sum_{G,G_L}\sum_{C_f}\sum_{C_S}\sum_{C_{\bar{P}}}
    \int \frac{dk^+_{\bar{P}}}{2\pi}
    \prod_l \int \frac{d^4 q_l}{(2\pi)^4}
    \nn \\ &\hspace{0.5cm}\times\;
    \Bigl[J^{(C_{\bar{P}})}_{\bar{P}}\Bigr]^{\mu_1\cdots\mu_n}
    \!\!\left(k^+_{\bar{P}}, \{q_l\}\right)
    \;S^{(C_S)\;\nu_1\cdots\nu_m}_{\mu_1\cdots\mu_n}\!\left(\{q_l\}\right)
    \nn \\ &\hspace{0.5cm}\times\;
    \prod_{j=1}^{N_j}
    \left[J_j\right]^{(C_f)}_{\nu_{l_j}\cdots}
    \!\!\left(k_j\cdot\bar{n}_j, \{q_l\}\right)
    \;H^{\rho\sigma}\!\left(\{k_j\}\right),
    \label{eq:Gamma2_region}
\end{align}
where, crucially, $H^{\rho\sigma}$ receives \emph{no} collinear line from $J_{\bar{P}}$.
Applying the Grammer--Yennie decomposition and the eikonal identity to the
final-state jets exactly as in Sec.~\ref{subsec:fact-residual}, and summing
inclusively over all cuts of the soft subgraph $S$, the KLN cancellation of
Eq.~\eqref{eq:soft_unity} applies, leaving
\begin{equation}
    \Gamma^{\rho\sigma,(2)}\big|_{L}
    \;=\;
    \sum_{C_{\bar{P}}}
    \int\frac{dk^+_{\bar{P}}}{2\pi}
    \Bigl[J^{(C_{\bar{P}})}_{\bar{P},\,\mathrm{SE}}\Bigr]\!\left(k^+_{\bar{P}}\right)
    \;\times\;
    \prod_{j=1}^{N_j}\int\frac{d(k_j\cdot\bar{n}_j)}{2\pi}
    \Bigl[J^{(C_f)}_j(k_j\cdot\bar{n}_j)\Bigr]
    \;H^{\rho\sigma}\!\left(\{k_j\}\right).
    \label{eq:Gamma2_SE}
\end{equation}
The result is a fully disconnected $\bar{P}$-jet: a partonic self-energy with $\bar{P}$
entering and exiting. The self-energy jet $J_{\bar{P},\,\mathrm{SE}}$ supplies only a
single pole in $\bar{P}^2 - m^2$, whereas LSZ reduction on the external $\bar{P}$ leg
requires a double pole. Therefore
\begin{equation}
    \lim_{\bar{P}^2\to m^2}(\bar{P}^2 - m^2)^2\;
    \Bigl[J^{(C_{\bar{P}})}_{\bar{P},\,\mathrm{SE}}\Bigr]
    \;=\; 0,
\end{equation}
and the connected contribution to $\Gamma^{\mu\nu,(2)}$ vanishes identically. The
IR divergences of $\Gamma^{\mu\nu,(2)}$ therefore reside entirely in the partially
disconnected graphs, to which we now turn.

\paragraph{Factorization of the partially disconnected contribution to 
$\Gamma^{\mu\nu,(2)}$.}
Having established that the connected contribution is IR finite, we now turn to the
partially disconnected graphs, which carry all the IR divergences of
$\Gamma^{\mu\nu,(2)}$. The power counting argument is as follows: at leading power,
each jet subgraph may emit exactly one physical parton into the hard function, with
any additional partons required to be longitudinal and accounted for by Wilson lines.
In the partially disconnected graphs, the amplitude jet emits one physical parton with
momentum $\xi\bar{P}$ into the hard function. A second independent physical parton,
were it to remain on the amplitude side, would be power suppressed; it is therefore
forced to escape to the c.c. amplitude, where it is the single physical parton emitted
by the c.c. jet at leading power. The contribution from region $L$ near the leading
pinch takes the form
\begin{align}
    \Gamma^{\rho\sigma,(2)}\big|_L
    &=
    \sum_{G,G_L}\sum_{C_f}\sum_{C_S}\sum_{C_{\bar{P}}}
    \int \frac{dk^+_{\bar{P}}}{2\pi}
    \prod_l \int \frac{d^4 q_l}{(2\pi)^4}
    \nn \\ &\hspace{0.5cm}\times\;
    \Bigl[J^{(C_{\bar{P}})}_{\bar{P}}\Bigr]^{\mu_1\cdots\mu_n}
    \!\!\left(k^+_{\bar{P}}, \{q_l\}\right)
    \;S^{(C_S)\;\nu_1\cdots\nu_m}_{\mu_1\cdots\mu_n}\!\left(\{q_l\}\right)
    \nn \\ &\hspace{0.5cm}\times\;
    \prod_{j=1}^{N_j}
    \left[J_j\right]^{(C_f)}_{\nu_{l_j}\cdots}
    \!\!\left(k_j\cdot\bar{n}_j, \{q_l\}\right)
    \;H^{\rho}\!\left(k^+_{\bar{P}},\{k_j\}\right)
    \;H^{*\sigma}\!\left(\{k_j\}\right)
    \;+\;\text{c.c.},
    \label{eq:Gamma2_disconn_region}
\end{align}
where the c.c.\ exchanges the roles of $H^\rho$ and $H^{*\sigma}$, and the absence
of $k^+_{\bar{P}}$ in $H^{*\sigma}$ reflects the fact that the $\bar{P}$-jet does
not enter the c.c.\ hard function. Applying the Grammer--Yennie decomposition and
eikonal identity to the final-state jets and summing inclusively over all soft cuts,
the KLN cancellation of Eq.~\eqref{eq:soft_unity} applies unchanged,
\begin{equation}
    \sum_{C_S}
    \prod_j \mathcal{W}_{n_j}\bigl[\{q_l\}\bigr]_{\nu_{l_j}\cdots}
    S^{\nu_1\cdots\nu_m}_{\mu_1\cdots\mu_n}(\{q_l\})
    \Bigl[J^{(C_{\bar{P}})}_{\bar{P}}\Bigr]^{\mu_1\cdots\mu_n}
    \!\!\left(k^+_{\bar{P}}, \{q_l\}\right)
    \;=\; \mathbf{1}
    \Bigl[J^{(C_{\bar{P}})}_{\bar{P}}\Bigr]\!\left(k^+_{\bar{P}}\right),
\end{equation}
leaving the residual $\bar{P}$-collinear jet depending only on $k^+_{\bar{P}}$.
The discontinuity of this jet, summed over cuts, is identified with the DIS PDF
exactly as in Eq.~\eqref{eq:jet_is_PDF},
\begin{equation}
    \sum_{C_{\bar{P}}}
    \bigl[J^{(C_{\bar{P}})}_{\bar{P}}\bigr](\xi P^+)
    \;=\;
    \int\frac{d\lambda}{2\pi}e^{-i\lambda\xi\bar{P}\cdot n}
    \langle\bar{P}|\,\bar{\psi}(\lambda n)\,\gamma^+\,
    \mathcal{W}_{n^-}(\lambda n, 0)\,\psi(0)\,|\bar{P}\rangle
    \;\equiv\; f_i(\xi,\mu_r^2),
\end{equation}
in exact correspondence with Eq.~\eqref{eq:jet_is_PDF}. Projecting onto scalar
structures and assembling over all leading regions, including the conjugate
configuration in which the roles of $H^\rho$ and $H^{*\sigma}$ are exchanged,
then yields
\begin{equation}
    \Gamma^{(2)}_a\left(q^2, x_F=\frac{1}{\o_F}\right)
    \;=\;
    \sum_i\int_0^1\frac{d\xi}{\xi}\;
    K^{(2)}_{a,i}\!\left(\frac{\o_F}{\xi},\frac{q^2}{\mu_r^2}\right)
    f_i(\xi,\mu_r^2)
    \;+\;\mathcal{O}\!\left(\frac{1}{Q^2}\right),
    \label{eq:Gamma2_factorized}
\end{equation}
where $K^{(2)}_{a,i}$ is a new perturbative coefficient computable order by order in
$\alpha_s$. Combining Eqs.~\eqref{eq:Gamma1_factorized} and
\eqref{eq:Gamma2_factorized} establishes Eq.~\eqref{eq:disc_factorization} for the
full residual function $\Gamma_a = \Gamma^{(1)}_a + \Gamma^{(2)}_a$.
\paragraph{Result.}
Assembling Eqs.~\eqref{eq:Gamma1_factorized}--(\ref{eq:Gamma2_factorized}) and performing
the convolution in $\x$, we obtain
\be
  \G_a\left(q^2, x_F=\frac{1}{\o_F}\right)
  \;=\;
  \sum_i \int_0^1 \frac{d\x}{\x}\;
  K_{a,i}\!\left(\frac{\o_F}{\x},\frac{Q^2}{\m_r^2}\right)
  f_i(\x,\m_r^2)
  \;+\; \mathcal{O}\!\left(\frac{1}{Q^2}\right),
  \label{eq:Gamma-final}
\ee
which establishes Eq.~\eqref{eq:disc_factorization}.  The only
long-distance input is $f_i(\x,\m_r^2)$, identical to the parton
distribution appearing in the DIS factorization theorem
Eq.~\eqref{eq:DIS_factorization}: no new non-perturbative data is
required to describe $\G_a$ at leading power.  The coefficient
$K_{a,i}$ is a new perturbative object, computable order by order in
$\a_s$ but distinct from the DIS coefficient $C_{a,i}$. It is clear that $K_i^{\mu\nu}$ is to be identified with the IR finite part of the partonic cross section $\hat{\Gamma}^{\mu\nu}(\xi\bar{P},q)$ which is to be understood as the partonic calculation of $\Gamma^{\mu\n}$ as defined in Eqs. (\ref{eq:gamma-decomp}),(\ref{eq:gamma-algorithm}).

\begin{figure}[ht]
  \centering
  \includegraphics[]{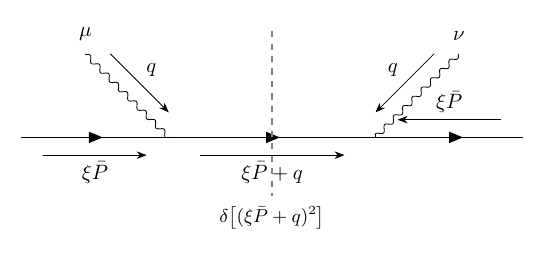}
  \caption{Leading-order contribution to the residual tensor $\Gamma^{\mu\nu}$ (and $\G^{\m\n,(1)}$)
           in the parton model. The incoming parton carries momentum $\xi\bar{P}$
           and the virtual photon carries momentum $q$. The dashed line denotes
           the final-state cut, with the on-shell condition
           $\delta[(\xi\bar{P}+q)^2]$ enforcing $q^2=0$.}
  \label{fig:dis_lo}
\end{figure}

Finally, let us discuss the leading order calculation of the hard coefficient $K_{a,i}$. Cuts of type (ii) do not appear at leading order, and $\G^{\m\n}$ is entirely made of cuts of type (iii). The graph is identical to the leading order DIS hadronic tensor graph and is shown in Fig.\ \ref{fig:dis_lo}.
While the Feynman rules are identical to the lowest order DIS calculation, the momentum conservation conservation is distinct in the $q^2>0$ region. The result reads
\bea
\hat{\Gamma}^{\m\n}(\xi\bar{P},q)&=&\frac{N_fN_c}{8\pi}\sum_{s_1,s} \int \frac{d^3l}{(2\pi)^32\o_l} (\bar{v}_{s_1}(\xi \bar{P})\gamma^{\m}v_s(l)) (2\pi)^4\d^4(\bar{P}+q-l)\nn \\ 
&&  \hspace{2cm}\times (\bar{v}_{s}(l)\gamma^{\n}v_{s_1}(\xi \bar{P})),\nn \\ 
&=&  N_fN_c\theta(q^0+P^0) \delta(q^2) \frac{\xi}{\xi x_F+1} \left(\bar{P}^{\m}l^{\n}+\bar{P}^{\n}l^{\m}-\eta^{\m\n}(\bar{P}\cdot l)\right). 
\eea
At this order, there are no initial state singularities and the hard coefficients are projected by the relationship
\bea
\hat{\Gamma}^{\m\n}(\xi\bar{P},q)&=&\left(\frac{q^{\m}q^{\n}}{q^2}-\eta^{\m\n}\right)
       K_{1}(q^2,2\xi\bar{P}\cdot q)
  \nn \\ &&
  -\frac{x_F^2}{4}
       \left(\frac{2}{x_F}\xi \bar{P}^\m-q^{\m}\right)
       \left(\frac{2}{x_F}\xi \bar{P}^\n-q^{\n}\right)
       \frac{K_2 (q^2,2 \xi \bar{P}\cdot q)}{\bar{P}\cdot q},
\eea
which is only valid at leading order.
Reading off the scalar hard kernals we get,
\bea
K_{1}&=&  -N_fN_c \theta(q^0+P^0) \delta(q^2) \frac{x_F q^2}{\xi x_F+1} \nn \\ 
K_{2}&=&  N_fN_c \theta(q^0+P^0) \delta(q^2) \frac{x_F q^2}{2(\xi x_F+1)},
\eea
which satisfies the Bjorken's scaling relation $K_2=-2K_1$ in the crossed region. Somewhat miraculously, both the kernels are zero since they are both of the form $q^2\d(q^2)$. At lowest order, they are not supported anywhere over  the  domain of the dispersive integral. As a result, we may drop the $\widehat{K} $ term in Eq.\ (\ref{eq:master_moment}) in the LO approximation. It then becomes evident that Eq.\ (\ref{eq:master_moment})  is a Gribov-Lipatov like relationship between Mellin moments of distribution functions and fragmentation functions. 
This concludes our lowest order calculation of the hard kernel. Starting at NLO, gluon initiated time-like DIS has a potential non-vanishing contribution to $\hat{\G}^{\m\n,(2)}$ while the quark initiated processes continue to be familiar, DIS like contributions to $\G^{\m\n,(1)}$. It would be particularly interesting to compute the NLO kernel in order to understand the role of small $q^2$ region in the dispersion relation Eq.\ (\ref{eq:dispersion-SIA}). Furthermore the question of wether  the residual kernels $K_a$ are non-zero at NLO remains interesting.  
\section{Conclusions}
\label{sec:conclusions}
In this work we have derived a dispersive crossing relation connecting
spacelike and timelike inclusive structure functions.  Starting from
the observation of Drell, Levy, and Yan that the DIS and SIA structure
functions are discontinuities of a single analytic forward amplitude, up to residual double discontinuities, 
and supplementing this with polynomial boundedness, we derived
twice-subtracted dispersion relations that relate the two experimentally
accessible regions without any model input.  The analyticity
assumptions underlying these dispersion relations were justified to all
orders in perturbation theory via an explicit analysis of the
Feynman-parametric representation of the amplitude, using the
Symanzik polynomial structure and combinatorial arguments . We also demonstrated that the residual double discontinuities factorizes within the partonic approximation, with the long distance matrix element being identical to the DIS matrix element. The short distance function however, is a new, perturbative object which can be computed order by order. 

At leading power, collinear factorization on both sides of the
dispersion relation transforms it into a concrete identity: a
convolution of DIS coefficient functions against parton distribution
functions on the spacelike side equals a sum of Mellin moments of
fragmentation functions weighted by dispersion relation transformed SIA hard coefficients on
the timelike side. This was a direct consequence of the absence of new non-perturbative data in residual double discontinuity function.  This opens a practical avenue: given sufficiently
precise DIS structure function data across a wide range of $x$ and
$Q^2$, one may in principle extract the Mellin moments
$\tilde{D}_i(N,\mu_D^2)$ of fragmentation functions directly, without
independent $e^+e^-$ measurements.  The dispersive relation Eq.\ (\ref{eq:residual-mellin}) differs qualitatively from its DIS counterpart in that the residual function $\Gamma_a$ receives contributions from arbitrarily small parton momentum fractions $\xi$, due to the absence of the spectral constraint that enforces $\xi >x_B$ in the spacelike case. This has two consequences. First, the practical usefulness of the dispersive integral depends on the contribution of the PDF and the hard kernel $K_{a,i}$ at small $\xi$, which requires further study, particularly at NLO. Of course, by choosing suitably large subtractions point $Q_i^2$, the effects of the NLO hard coefficient of the residual function may be reduced arbitrarily.  Second, the relation is sensitive to the small-x PDF in a way that standard DIS at moderate Bjorken x is not — suggesting a potential phenomenological application in constraining small-x parton distributions from timelike inclusive data.

More broadly, the dispersive framework established here connects two
sets of intrinsically infrared quantities — parton distributions and
fragmentation functions — through a relation whose derivation is
controlled entirely by analyticity.  Relating
nonperturbative infrared data on both sides of the dispersion via the
analytic structure of the hard amplitude is, we believe, an
interesting direction that deserves further exploration, both
perturbatively and beyond.

\section*{Acknowledgements}
I would like to thank Henry Klest for enthusiasm and conversation during the early stages of this work.
I would also like to thank George Sterman for a helpful exchange regarding known obstructions to crossing.

\appendix
\section{Analyticity in Perturbation Theory}
\label{app:PT-analyticity}

This appendix proves that the amplitudes $T_a$ satisfy the cut-plane analyticity stated in Assumption 1 of Sec.~\ref{sec:field-theory}, to all orders in perturbation theory.
The proof has two components: analyticity in the complex
$x$-plane for fixed spacelike $Q^2$, and analyticity in the complex
$Q^2$-plane for fixed $x > 1$.

Our strategy to prove analyticity in the complex $x$-plane is to first
establish analyticity in the strip $\bigl|\Re(1/x)\bigr|<1$,
demonstrate reality on the real axis, and finally establish analyticity
for $\bigl|\Im(1/x)\bigr|\neq 0$ with $\bigl|\Re(1/x)\bigr|>1$.
This will establish analyticity everywhere in the cut plane at fixed
$Q^2\leq 0$. In contrast, we will establish analyticity in the cut
$Q^2$-plane by first establishing analyticity in the
$\Im(Q^2)>0$ half-plane at fixed $x>1$, and reality on the negative
real $Q^2$ axis.  The continuation to the $\Im(Q^2)<0$ half-plane is
then fixed by the Schwartz reflection principle.

In perturbation theory, the scalar invariant amplitudes $T_a$ receive
contributions from individual Feynman diagrams.  Let $\mathcal{G}$
denote the set of all Feynman graphs contributing to the forward
virtual Compton scattering amplitude at any given loop order.  A
generic graph $G\in\mathcal{G}$ has $L$ independent loop momenta
$\ell_1,\ldots,\ell_L$, $E$ internal propagators with momenta
$k_1,\ldots,k_E$ (each a linear combination of the loop momenta and
the external momenta $P$ and $q$), and carries a Lorentz- and
spinor-index numerator that we denote symbolically by $\mathcal{N}_G$.
The amplitude is therefore a sum over graphs,
\begin{equation}
  T_a \;=\; \sum_{G\,\in\,\mathcal{G}}\;
  \int \prod_{i=1}^{L} \frac{d^4\ell_i}{(2\pi)^4}\;
  \frac{\mathcal{N}_G\!\left(\ell_1,\ldots,\ell_L;\,P,q\right)}
       {\displaystyle\prod_{e=1}^{E}\bigl(k_e^2 - m_e^2 + i\varepsilon\bigr)},
  \label{eq:T_a_Feynman}
\end{equation}
where $m_e$ is the mass of the particle on propagator $e$, and the
$+i\varepsilon$ prescription ($\varepsilon>0$, to be taken to zero at
the end) encodes the standard causal time-ordering of the Feynman
propagator.  The numerator $\mathcal{N}_G$ is a polynomial in all
momenta and, as we discuss momentarily, does not introduce any
additional singularities in the external kinematics.

\subsubsection*{Feynman parametrization}

To analyse the analytic structure of \eqref{eq:T_a_Feynman}, we
combine the $E$ propagator denominators using the Feynman
parametrization formula,
\begin{equation}
  \prod_{e=1}^{E} \frac{1}{D_e}
  \;=\;
  (E-1)!\int_0^1 \prod_{e=1}^{E} d\alpha_e\;
  \delta\!\left(1 - \sum_{e=1}^{E}\alpha_e\right)
  \frac{1}{\left(\displaystyle\sum_{e=1}^{E}\alpha_e D_e\right)^{E}},
  \label{eq:Feynman_param_formula}
\end{equation}
where $D_e = k_e^2 - m_e^2 + i\varepsilon$.  After introducing the
Feynman parameters $\{\alpha_e\}$ and performing a standard shift of
the loop integration variables to complete the square, the combined
denominator takes the form
\begin{equation}
  \mathcal{D}_G
  \;=\;
  \sum_{e=1}^{E}\alpha_e\,D_e
  \;=\;
  \sum_{e=1}^{E}\alpha_e\bigl(k_e^2 - m_e^2\bigr)
  + i\varepsilon.
  \label{eq:Feynman_denom_symbolic}
\end{equation}
The single $i\varepsilon$ here arises from summing the individual
$i\varepsilon$ prescriptions with positive weights $\alpha_e \geq 0$;
crucially, the sign of the imaginary part is preserved throughout.

\subsubsection*{Numerator analyticity}

Before examining the denominator in detail, let us note that the
numerator $\mathcal{N}_G$ is automatically analytic in the external
momenta.  After the loop-momentum shift required to complete the
square in the denominator, the numerator becomes a polynomial in the
shifted loop momenta (which are dummy integration variables) and in the
external kinematic invariants.  Polynomial functions are entire; they
introduce no branch cuts and no poles.  Consequently, \emph{all}
singularities of a given Feynman diagram in the external kinematic
variables $Q^2$ and $x$ reside entirely in the denominator
$\mathcal{D}$.  The analyticity question therefore reduces entirely to
the analysis of the parametric denominator.

\subsubsection*{The parametric denominator in terms of loop momenta}

It remains to make the quadratic structure of $\mathcal{D}$ in the
loop momenta explicit.  Since each internal momentum $k_e$ is a linear
function of the loop momenta $\ell_i$ and the external momenta $q,P$, we may write
\begin{equation}
  \mathcal{D}_G
  \;=\;
  \sum_{i,j=1}^{L} M_{ij}(\alpha)\,\ell_i \cdot \ell_j
  \;-\; 2\sum_{i=1}^{L} \ell_i \cdot q_i(\alpha, q, P)
  \;+\; J(\alpha, q, P, m^2)
  \;+\; i\varepsilon ,
  \label{eq:Denom_loop_form}
\end{equation}
where the quantities appearing here are:
\begin{itemize}
  \item $M_{ij}(\alpha)$ is a symmetric, positive-semidefinite
    $L\times L$ matrix whose entries are linear in the Feynman
    parameters $\{\alpha_e\}$; it encodes the combinatorics of which
    loop momenta flow through each propagator.  Positive semidefiniteness
    of $M_{ij}$ on the physical sheet is a consequence of the
    $\alpha_e \geq 0$ constraints.
  \item $q_i(\alpha, q, P)$ is a linear combination of external momenta
    with coefficients that are linear in $\{\alpha_e\}$; the factor of
    $2$ is the conventional normalisation that allows one to complete
    the square by the shift $\ell_i \to \ell_i - (M^{-1})_{ij}q_j$.
 \item $J(\alpha, q, P, m^2)$ is the term in $\mathcal{D}$ that is
    independent of the loop momenta; it is linear in the Feynman
    parameters $\{\alpha_e\}$ and encodes the dependence on the external
    kinematics and internal masses.  Schematically,
    \begin{equation}
      J(\alpha, q, P, m^2)
      \;=\;
      \sum_{e=1}^{E} \alpha_e\,\bigl(r_e(q,P)^2 - m_e^2\bigr),
    \end{equation}
    where $r_e(q,P)$ denotes the purely external-momentum part of $k_e$
    (i.e.\ the value of the internal momentum on line $e$ when all loop
    momenta are set to zero).  The dependence on the external invariants
    $Q^2$ and $x$ therefore enters $\mathcal{D}$ exclusively through
    $J$ and $q$.
  \item The $+i\varepsilon$ in \eqref{eq:Denom_loop_form} is the
    ubiquitous Feynman prescription, retained explicitly throughout.
\end{itemize}
We now perform the shift of loop momenta that eliminates the linear
term in \eqref{eq:Denom_loop_form} and carry out the $l_i$ integrals.  Since $M_{ij}$ is invertible on
the interior of the simplex, we first shift
$\ell_i'= \ell_i - (M^{-1})_{ij}\,q_j$,
under which the denominator becomes
\begin{equation}
  \mathcal{D}_G
  =
  \sum_{i,j=1}^{L} M_{ij}(\alpha)\,\ell'_i \cdot \ell'_j
  \;+\; \frac{\mathcal{F}_G(\alpha, q, P, m^2)}{ \mathcal{U}_G(\alpha)}
  \;+\; i\varepsilon,
  \label{eq:Denom_shifted}
\end{equation}
where the two Symanzik polynomials, $\mathcal{U}_G$ and
$\mathcal{F}_G$, have emerged naturally.  The Jacobian of the shift
is unity, but upon Wick rotating and carrying out the resulting Gaussian integrals, the rescaling of loop momenta produces a factor of $(\det M)^{-L/2}$ which may be absorbed in the measure of the $\a$ integrals; this factor is the first
Symanzik polynomial,
\begin{equation}
  \mathcal{U}_G(\alpha) \;=\; \det M_{ij}(\alpha).
  \label{eq:U_def}
\end{equation}
The shift also produces a new constant term in the denominator,
which combines with the original $J$,
\begin{equation}
  \frac{\mathcal{F}_G(\alpha, q, P, m^2)}{\mathcal{U}_G(\alpha)}
  \;=\;
  J(\alpha, q, P, m^2)
  \;-\;
  \sum_{i,j=1}^{L}(M^{-1})_{ij}(\alpha)\;q_i(\alpha,q,P)\cdot q_j(\alpha,q,P).
  \label{eq:F_def}
\end{equation}
$ {\mathcal{F}_G(\alpha, q, P, m^2)}$ is the second Symanzik polynomial.  Both $\mathcal{U}_G$ and
$\mathcal{F}_G$ are polynomials in the Feynman parameters
$\{\alpha_e\}$ with non-negative integer coefficients, a fact made
transparent by their graph-theoretic definitions, to which we now
turn.

\subsubsection*{Spanning trees and the graph-theoretic representation}

A \emph{spanning tree} of $G$ is a subgraph that contains every vertex
of $G$, is connected, and contains no closed loops; it is the minimal
set of edges that keeps the graph in one piece.  A graph with $V$
vertices and $E$ edges has spanning trees each consisting of exactly
$V-1$ edges, so that removing any one of the remaining $E-(V-1)$ edges
from a spanning tree disconnects it.  A \emph{spanning 2-tree} of $G$
is precisely such a disconnected subgraph: it is obtained from a
spanning tree by deleting one of its edges, leaving a pair of connected,
acyclic components $T^{(2)}_1$ and $T^{(2)}_2$ that together still
cover every vertex of $G$.  We denote by $P_1$ the sum of all external
momenta entering the component $T^{(2)}_1$; momentum conservation then
forces $P_2 = -P_1$ for the other component.

The first Symanzik polynomial $\mathcal{U}_G$ counts spanning trees,
weighted by the product of the Feynman parameters on the edges
\emph{not} in the tree,
\begin{equation}
  \mathcal{U}_G(\alpha)
  \;=\;
  \sum_{T\,\in\,\mathcal{T}_G}
  \;\prod_{e\,\notin\, T} \alpha_e,
  \label{eq:U_spanning_tree}
\end{equation}
where the sum runs over all spanning trees $T$ of $G$ and the product
runs over the edges of $G$ absent from $T$.  Because every term in
this sum is a product of non-negative $\alpha_e$'s,
$\mathcal{U}_G \geq 0$ on the integration simplex, with equality only
at the boundary where some $\alpha_e \to 0$.

The second Symanzik polynomial $\mathcal{F}_G$ similarly counts
spanning 2-trees, but now each 2-tree carries an additional kinematic
weight from the external momentum flowing between its two components,
\begin{equation}
  \mathcal{F}_G(\alpha, q, P, m^2)
  \;=\;
  \sum_{T^{(2)}\,\in\,\mathcal{T}^{(2)}_G}
  P_1^2\prod_{e\,\notin\, T^{(2)}} \alpha_e
  \;-\;
  \mathcal{U}_G(\alpha)\sum_{e=1}^{E}\alpha_e m_e^2,
  \label{eq:F_spanning_2tree}
\end{equation}
where the sum runs over all spanning 2-trees $T^{(2)}$ of $G$,
$P_1$ is the total external momentum entering one of the two
components, and the product again runs over edges absent from $T^{(2)}$.
The second term, proportional to $\mathcal{U}_G$, accounts for the
internal masses.  

The forward Compton amplitude has four external legs carrying momenta
$P,\,-P,\,q,\,-q$.  A spanning 2-tree $T^{(2)}$ partitions these four
legs between its two components; the distinct values of $P_1^2$ are
therefore $m_p^2$, $(P+q)^2$, $(P-q)^2$, and $q^2$.
Separating the kinematic sum accordingly, and isolating the proton
mass from the internal mass sum, we define
\begin{align}
  \mathcal{F}^{(P)}_G(\alpha) &\;=\;
  \sum_{\substack{T^{(2)}:\\ P_1 = P}}\prod_{e\notin T^{(2)}}\!\alpha_e,
  &
  \mathcal{F}^{(P+q)}_G(\alpha) &\;=\;
  \sum_{\substack{T^{(2)}:\\ P_1 = P+q}}\prod_{e\notin T^{(2)}}\!\alpha_e,
  \nonumber\\
  \mathcal{F}^{(P-q)}_G(\alpha) &\;=\;
  \sum_{\substack{T^{(2)}:\\ P_1 = P-q}}\prod_{e\notin T^{(2)}}\!\alpha_e,
  &
  \mathcal{F}^{(q)}_G(\alpha) &\;=\;
  \sum_{\substack{T^{(2)}:\\ P_1 = q}}\prod_{e\notin T^{(2)}}\!\alpha_e,
  \label{eq:F_pieces}
\end{align}
so that $\mathcal{F}_G$ decomposes as
\bea
  \mathcal{F}_G(\alpha,q,P,m^2)
  \;&=&\;
  m_p^2\,\mathcal{F}^{(P)}_G
  \;+\; (P{+}q)^2\,\mathcal{F}^{(P+q)}_G
  \;+\; (P{-}q)^2\,\mathcal{F}^{(P-q)}_G
  \;-\; Q^2\,\mathcal{F}^{(q)}_G \nn \\ &&
  \;-\; \mathcal{U}_G\sum_{e:\,m_e \neq m_p}\!\alpha_e m_e^2
  \;-\; \mathcal{U}_G\,m_p^2\!\sum_{e\in\mathcal{F}}\alpha_e.
  \label{eq:F_decomposed}
\eea
The non-proton mass terms are manifestly non-positive since
$\mathcal{U}_G,\alpha_e \geq 0$.  The only sources of $m_p^2$ are
the first and last terms; we study their combined sign next.

\subsubsection*{The coefficient of $m_p^2$ in $\mathcal{F}_G$ is non-positive}

Every $G\in\mathcal{G}$ contains an open fermion line $\mathcal{F}$,
a path of edges from the vertex at $P$ to the vertex at $-P$.
For any spanning 2-tree $T^{(2)}$ with components $T^{(2)}_1\ni P$,
$T^{(2)}_2 \ni {-P}$, there exists a unique edge $j$ with
\begin{equation}
  j \in \mathcal{F}, \qquad j \notin T^{(2)}, \qquad
  \partial j \cap T^{(2)}_1 \neq \emptyset, \qquad
  \partial j \cap T^{(2)}_2 \neq \emptyset,
  \label{eq:j_conditions}
\end{equation}
where $\partial j$ denotes the two endpoints of $j$.  The map
\begin{equation}
  \phi\colon
  \bigl\{\,T^{(2)} \in \mathcal{T}^{(2)}_G : P\in T^{(2)}_1,\,-P\in T^{(2)}_2\,\bigr\}
  \;\hookrightarrow\;
  \mathcal{T}_G,
  \qquad T^{(2)} \mapsto T^{(2)} \cup \{j\}
  \label{eq:injection}
\end{equation}
is an injection into the set of spanning trees of $G$.  The
contribution of a paired $(T^{(2)}, T = \phi(T^{(2)}))$ to the
coefficient of $m_p^2$ in $\mathcal{F}_G$ is
\begin{equation}
  \underbrace{m_p^2 \prod_{e\notin T^{(2)}}\alpha_e}_{\text{from }T^{(2)},\; P_1^2=m_p^2}
  \;-\;
  \underbrace{m_p^2\,\alpha_j\prod_{e\notin T}\alpha_e}_{\text{from }T,\;\text{mass term}}
  \;=\;
  m_p^2\prod_{e\notin T^{(2)}}\alpha_e
  \;-\;
  m_p^2\prod_{e\notin T^{(2)}}\alpha_e
  \;=\; 0,
  \label{eq:cancellation}
\end{equation}
where we used $\prod_{e\notin T}\alpha_e = \alpha_j^{-1}\prod_{e\notin T^{(2)}}\alpha_e$.
Spanning trees outside the image of $\phi$ contribute only
$-m_e^2(\cdots)\leq 0$ through the mass term. Hence
\begin{equation}
  \bigl[m_p^2\bigr]_{\mathcal{F}_G} \;\leq\; 0
  \qquad \forall\; \alpha_e \geq 0.
  \label{eq:mp2_nonpositive}
\end{equation}

\subsubsection*{Analyticity for $\bigl|\Re(x)\bigr|>1$ at fixed $q^2 < 0$}

With $q^2 < 0$ fixed and $x = -q^2/2P\cdot q$, the kinematic
invariants in \eqref{eq:F_decomposed} evaluate to
\begin{equation}
  (P+q)^2 \;=\; q^2\!\left(1 - \frac{1}{x}\right), \qquad
  (P-q)^2 \;=\; q^2\!\left(1 + \frac{1}{x}\right), \qquad
  -Q^2 \;=\; q^2.
  \label{eq:kinematics_x}
\end{equation}
Substituting into \eqref{eq:F_decomposed},
\begin{equation}
  \mathcal{F}_G \;=\; q^2\left[
    \left(1-\frac{1}{x}\right)\mathcal{F}^{(P+q)}_G
    \;+\;
    \left(1+\frac{1}{x}\right)\mathcal{F}^{(P-q)}_G
    \;+\;
    \mathcal{F}^{(q)}_G
  \right]
  \;+\; m_p^2\,\mathcal{F}^{(P)}_G
  \;-\; \mathcal{U}_G\sum_e \alpha_e m_e^2.
  \label{eq:F_in_x}
\end{equation}
For $\bigl|\Re(x)\bigr| > 1$ we have $\bigl|\Re(1/x)\bigr| < 1$, so
\begin{equation}
  \Re\!\left(1 - \frac{1}{x}\right) > 0, \qquad
  \Re\!\left(1 + \frac{1}{x}\right) > 0.
  \label{eq:Re_positive}
\end{equation}
Since $q^2 < 0$ and all $\mathcal{F}^{(\cdot)}_G \geq 0$, every term
in $\Re(\mathcal{F}_G)$ is non-positive:
\begin{align}
  q^2\,\Re\!\left(1\pm\frac{1}{x}\right)\mathcal{F}^{(P\pm q)}_G &\;\leq\; 0, \nonumber\\
  q^2\,\mathcal{F}^{(q)}_G &\;\leq\; 0, \nonumber\\
  m_p^2\,\mathcal{F}^{(P)}_G - \mathcal{U}_G\,m_p^2\!\sum_{e\in\mathcal{F}}\alpha_e &\;\leq\; 0, \nonumber\\
  -\mathcal{U}_G\sum_{e:\,m_e\neq m_p}\alpha_e m_e^2 &\;\leq\; 0.
  \label{eq:each_term_negative}
\end{align}
Therefore $\Re(\mathcal{F}_G) \leq 0$ for all $\alpha_e \geq 0$ when
$\bigl|\Re(x)\bigr|>1$.  Since $\mathcal{U}_G > 0$ on the interior of
the simplex, the denominator in \eqref{eq:Denom_shifted} satisfies
$\Re(\mathcal{F}_G/\mathcal{U}_G) \leq 0$, and after Wick rotation
$-\ell_i^2 + \mathcal{F}_G/\mathcal{U}_G + i\varepsilon$ never
vanishes.  No singularity of the Feynman integral is encountered along
any path in the region $\bigl|\Re(x)\bigr|>1$, establishing analyticity
of $T_a$ there.

\subsubsection*{Analytic continuation to $\bigl|\Re(x)\bigr| < 1$ at fixed $\Im(x)\neq 0$}

Consider a path at fixed $\Im(x) \neq 0$ starting from
$\bigl|\Re(x)\bigr|>1$ and moving into $\bigl|\Re(x)\bigr|<1$.
From \eqref{eq:F_in_x}, the imaginary part of $\mathcal{F}_G$ along
this path is
\begin{equation}
  \Im(\mathcal{F}_G)
  \;=\;
  q^2\,\Im\!\left(\frac{1}{x}\right)
  \bigl[
    \mathcal{F}^{(P-q)}_G - \mathcal{F}^{(P+q)}_G
  \bigr].
  \label{eq:ImF}
\end{equation}
$\Im(\mathcal{F}_G) = 0$ requires
\begin{equation}
  \mathcal{F}^{(P+q)}_G(\alpha) \;=\; \mathcal{F}^{(P-q)}_G(\alpha).
  \label{eq:ImF_zero_condition}
\end{equation}
On the subsurface $\Sigma$ defined by \eqref{eq:ImF_zero_condition},
substituting back into \eqref{eq:F_in_x} and using
$\Re(1-1/x) + \Re(1+1/x) = 2$,
\begin{equation}
  \Re(\mathcal{F}_G)\big|_{\Sigma}
  \;=\;
  2q^2\,\mathcal{F}^{(P+q)}_G
  \;+\; q^2\,\mathcal{F}^{(q)}_G
  \;+\; m_p^2\,\mathcal{F}^{(P)}_G
  \;-\; \mathcal{U}_G\sum_e \alpha_e m_e^2
  \;\leq\; 0.
  \label{eq:ReF_on_Sigma}
\end{equation}
$\mathcal{F}_G$ therefore cannot vanish along the path: vanishing
requires $\Im(\mathcal{F}_G) = \Re(\mathcal{F}_G) = 0$
simultaneously, but \eqref{eq:ReF_on_Sigma} is non-positive with
equality only at the boundary $\mathcal{U}_G = 0$ of the simplex.
$T_a$ is therefore analytic for all $\Im(x)\neq 0$ at fixed $q^2<0$.

\subsubsection*{Analytic continuation in $q^2$ at fixed real $x$ with $|x|>1$}

For real $x$ with $|x|>1$ and $q^2<0$, we have already established
that $\mathcal{F}_G/\mathcal{U}_G$ is real and non-positive, so the
denominator \eqref{eq:Denom_shifted} is real and non-vanishing.  We
now continue $q^2$ into the upper half-plane by writing
\begin{equation}
  q^2 \;=\; -|q^2|\,e^{-i\theta}, \qquad \theta \in (0,\pi).
  \label{eq:q2_continuation}
\end{equation}
Substituting into \eqref{eq:F_in_x}, the imaginary part of
$\mathcal{F}_G$ becomes
\begin{equation}
  \Im(\mathcal{F}_G)
  \;=\;
  |q^2|\sin\theta\left[
    \left(1-\frac{1}{x}\right)\mathcal{F}^{(P+q)}_G
    \;+\;
    \left(1+\frac{1}{x}\right)\mathcal{F}^{(P-q)}_G
    \;+\;
    \mathcal{F}^{(q)}_G
  \right]
  \;\geq\; 0,
  \label{eq:ImF_q2}
\end{equation}
since $\sin\theta > 0$ for $\theta\in(0,\pi)$, all
$\mathcal{F}^{(\cdot)}_G \geq 0$, and $(1\pm 1/x)>0$ for real
$|x|>1$.  The imaginary part \eqref{eq:ImF_q2} adds to the existing
$+i\varepsilon$, so $\Im(\mathcal{D}_G) > 0$ strictly throughout the
continuation.  The denominator is therefore non-vanishing at every
point along the path \eqref{eq:q2_continuation}, establishing
analyticity of $T_a$ in the upper half $q^2$-plane at fixed real
$x$ with $|x|>1$. 

We have established the analyticity properties of $T_a$ required for
the dispersion relations of Sec.~\ref{sec:results} entirely within
perturbation theory, graph by graph.  The argument rested on a single
structural fact — the Feynman-parametrized denominator
$\mathcal{F}_G/\mathcal{U}_G + i\varepsilon$ is non-vanishing on the
interior of the simplex in the relevant kinematic regions — which was
established by decomposing $\mathcal{F}_G$ into its four kinematic
pieces and the mass terms \eqref{eq:F_decomposed}, and analysing each
piece in turn.  Concretely:
\begin{itemize}
  \item The non-proton mass terms are manifestly non-positive.
  \item The proton mass terms cancel term-by-term via the open fermion
    line injection \eqref{eq:injection}, leaving a non-positive net
    contribution \eqref{eq:mp2_nonpositive}.
  \item For $|\Re(x)|>1$ at fixed $q^2<0$, every kinematic piece is
    non-positive \eqref{eq:each_term_negative}, so $\Re(\mathcal{F}_G)\leq 0$
    and the denominator is non-vanishing.
  \item Analyticity extends to all $\Im(x)\neq 0$ at fixed $q^2<0$:
    $\Im(\mathcal{F}_G)$ can vanish only on the subsurface $\Sigma$
    defined by \eqref{eq:ImF_zero_condition}, but on $\Sigma$ the real
    part \eqref{eq:ReF_on_Sigma} is strictly non-positive, so
    $\mathcal{F}_G\neq 0$ throughout.
  \item Analyticity in the upper half $q^2$-plane at fixed real
    $|x|>1$ follows from \eqref{eq:ImF_q2}: the imaginary part of
    $\mathcal{F}_G$ is positive definite for $\theta\in(0,\pi)$,
    reinforcing the $+i\varepsilon$ prescription.

    \end{itemize}

    This concludes our proof of analyticity used in the main text. 

\end{document}